\documentclass[final,3p,times]{elsarticle}

\usepackage{amssymb}
\usepackage{amsmath}
\usepackage{subcaption}
\usepackage{color}
\usepackage{glossaries}
\usepackage{hyperref}
\newacronym{lbm}{LBM}{lattice Boltzmann method}
\newacronym{weno}{WENO}{weighted-essentially-non-oscillatory}
\newacronym{sllbm}{SLLBM}{semi-Lagrangian lattice Boltzmann method}
\newacronym{dp}{DP}{departure point}
\newacronym{ap}{AP}{arrival point}
\newacronym{rms}{RMS}{root mean square}
\newacronym{dns}{DNS}{direct numerical simulation}

\journal{arXiv}

\begin{document}

\begin{frontmatter}



\title{Supersonic Shear and Wall-Bounded Flows With Body-Fitted Meshes Using the Semi-Lagrangian Lattice Boltzmann Method: Boundary Schemes and Applications}

\author[us,hbrs]{Philipp Spelten} \corref{cor1}
\author[us,hbrs]{Dominik Wilde}
\author[us,hbrs]{Mario Christopher Bedrunka}
\author[hbrs,scai]{Dirk Reith}
\author[us]{Holger Foysi}

\cortext[cor1]{Corresponding author: philipp.spelten@uni-siegen.de}

\address[us]{Department of Mechanical Engineering, Paul-Bonatz-Straße 9-11, 57076 Siegen-Weidenau, Germany}
\address[hbrs]{Institute of Technology, Resource and Energy-efficient Engineering (TREE),\\ Bonn-Rhein-Sieg University of Applied Sciences,
Grantham-Allee 20, 53757 Sankt Augustin, Germany}
\address[scai]{Fraunhofer Institute for Algorithms and Scientific Computing (SCAI), Schloss Birlinghoven, 53754 Sankt Augustin, Germany}

\begin{abstract}
Lattice Boltzmann method (LBM) simulations of incompressible flows are nowadays common and well-established. However, for compressible turbulent flows with strong variable density and intrinsic compressibility effects, results are relatively scarce. Only recently, progress was made regarding compressible LBM, usually applied to simple one and two-dimensional test cases due to the increased computational expense.
The recently developed semi-Lagrangian lattice Boltzmann method (SLLBM) is capable of simulating two- and three-dimensional viscous compressible flows. This paper presents bounce-back, thermal, inlet, and outlet boundary conditions new to the method and their application to problems including heated or cooled walls, often required for supersonic flow cases.
Using these boundary conditions, the SLLBM's capabilities are demonstrated in various test cases, including a supersonic 2D NACA-0012 airfoil, flow around a 3D sphere, and, to the best of our knowledge, for the first time, the 3D simulation of a supersonic turbulent channel flow at a bulk Mach number of $\mathrm{Ma}=1.5$ and a 3D temporal supersonic compressible mixing layer at convective Mach numbers ranging from $\mathrm{Ma}=0.3$ to $\mathrm{Ma}=1.2$. The results show that the compressible SLLBM is able to adequately capture intrinsic and variable density compressibility effects.
\end{abstract}

\begin{graphicalabstract}

\includegraphics[width=0.73\linewidth]{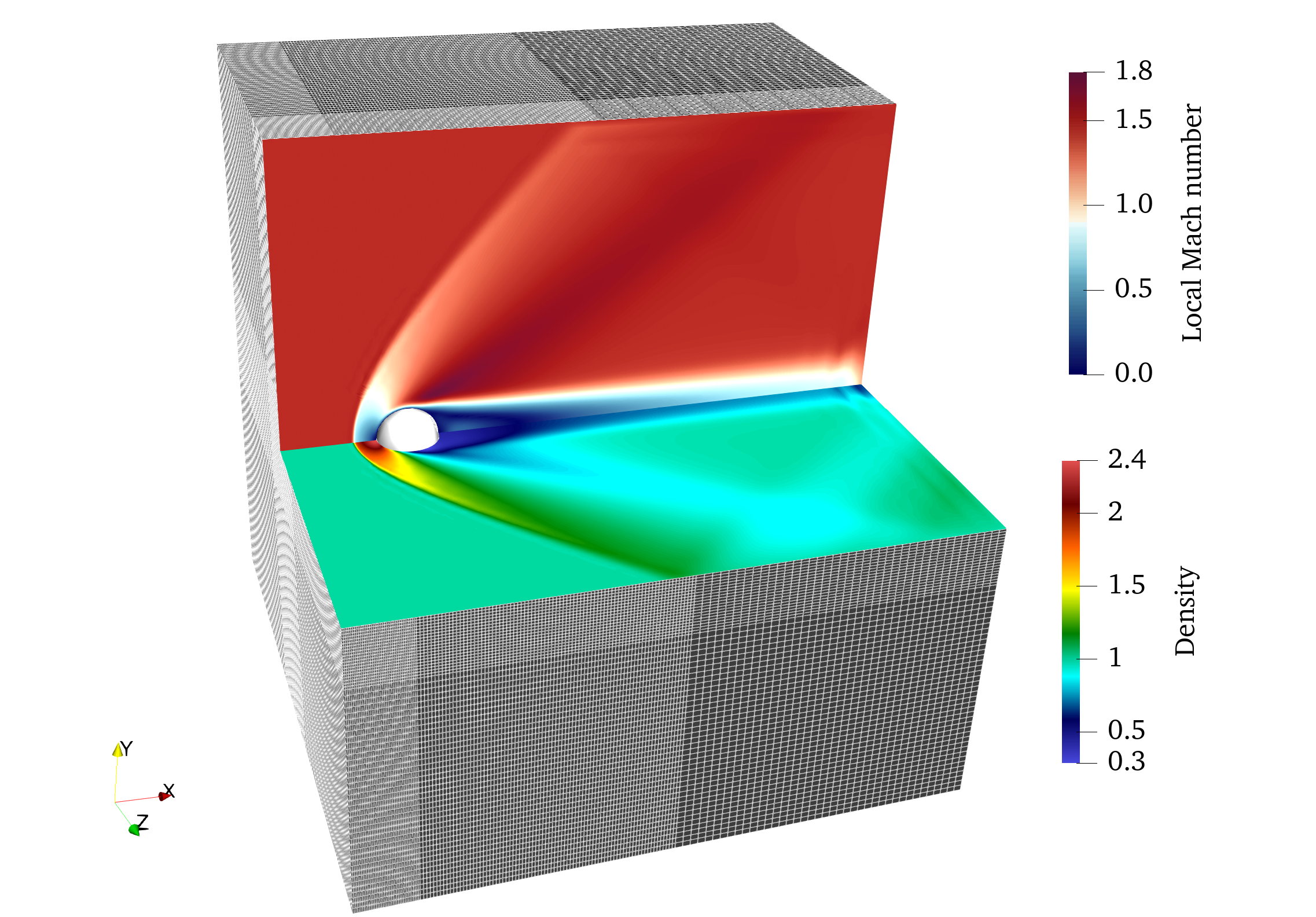}

\includegraphics[width=0.73\linewidth]{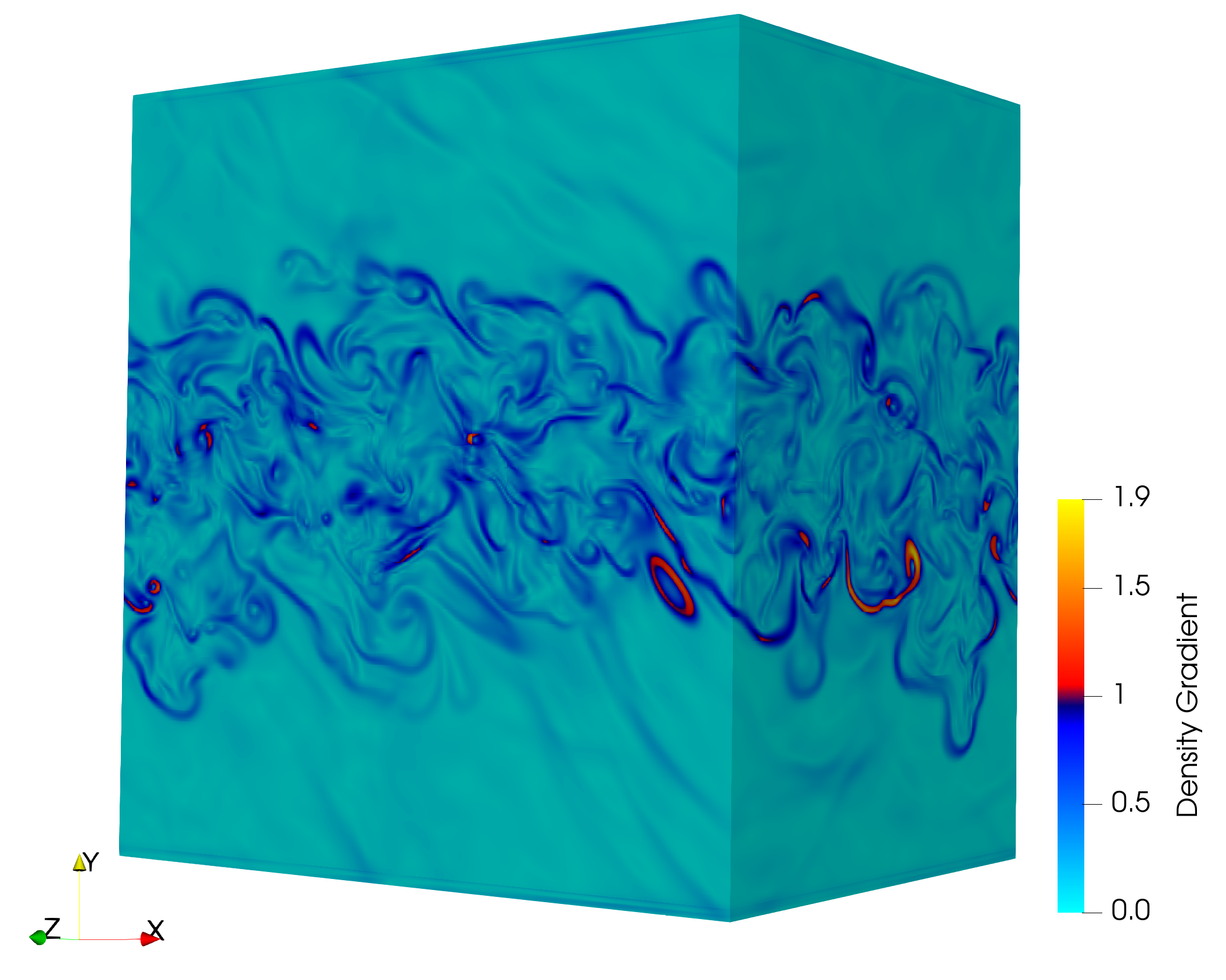}

\end{graphicalabstract}

\begin{highlights}
\item Adding temperature and equilibrium boundary conditions to the semi-Lagrangian lattice Boltzmann method (SLLBM).
\item The increased efficiency by employing cubature rules to derive a reduced velocity space is, for the first time, applied to 3D cooled flows and flows around objects.
\item Demonstrating the capabilities of the SLLBM for simulating the supersonic flow around a NACA-0012 airfoil in 2D and around a sphere in 3D.
\item For the first time, to the best of our knowledge, using the LBM to simulate the 3D supersonic flow in a turbulent channel and in a temporal compressible mixing layer.
\end{highlights}


\end{frontmatter}


\section{Introduction}
Viscous turbulent compressible subsonic and supersonic flows can appear in a variety of natural and human-made systems, such as in aerospace engineering, where high-speed airflows occur around aircraft and rockets, in rocket engines or in geophysical fluid dynamics, where volcanic flows or meteor impacts can generate highly compressible or even supersonic flows with significant variable property changes \cite{Kiefer:1981,Tam:1995,Oestlund:2005,Gatskibook,Orescanin:2014,Silber:2018}.

From a computational perspective, the simulation of viscous supersonic flows is a challenging task that requires advanced numerical methods and computational resources. The flow dynamics are influenced by a range of physical processes, including turbulence, shock waves, or heat transfer, making it difficult to accurately capture these flows using traditional numerical approaches \cite{Gatskibook,Carpenter:1990,Shu:2009,Qu:2022}.

Although advances in computational techniques and hardware have led to significant progress in the simulation of viscous supersonic flows and have opened up new opportunities for research in this field, numerical simulations of viscous supersonic flows present several challenges. First, supersonic flows often involve complex geometries such as those found in aerospace applications, which require suited boundary conditions in simulations. Second, supersonic flows around objects usually provoke shocks with local sharp gradients of the variables, which demands schemes with shock-capturing properties. Third, turbulent features and the turbulent interaction with shocks of supersonic flows must also be captured, despite the high Reynolds numbers and small timescales.

In the past, most works used solvers based on the Navier-Stokes equations, whereas methods based on the Boltzmann equations are more exotic.
Common numerical additions to solvers of the Navier-Stokes equations are types of artificial diffusivity via extra terms added to the governing equations \cite{Fiorina:2007,Kawai:2008,Ranocha:2018,Haga:2019} or selective filtering \cite{Bogey:2008,Bogey:2009,Patel:2019} and \gls{weno} schemes \cite{Shu:2009,Martin:2006,Sun:2021}. Ideally, these tools are minimally numerically invasive to allow for realistic laminar-turbulence transition and to obey conservation properties without excessive dissipation. Furthermore, they correctly predict shock-turbulence interaction in, for example, capturing the almost instant amplification in the streamwise vorticity variance behind the shock and post-shock Reynolds stress anisotropy \cite{Larsson:2009}.

The \gls{lbm} is a rapidly emerging alternative scheme to simulate, mostly weakly compressible, fluid behavior.
However, there are hardly any results shown for \gls{lbm} in three dimensions for compressible flows, which is mainly due to high demands in terms of computational costs and solver capabilities. Particularly shocks near the objects need special treatment. In the last decade, however, several works have extended the method's capability to compressible flow computations.

The \gls{sllbm} was developed by Krämer \emph{et al.} \cite{Kraemer:2017} and enhanced by Wilde \emph{et al.} \cite{Wilde2020}. This paper further augments the method to simulate three-dimensional, fully compressible, wall-bounded flows in Sec.~\ref{sec:method}.
In Sec.~\ref{sec:results}, we present a collection of results achieved using this method, demonstrating its capabilities. A summary and directions for further research are given in Sec.~\ref{sec:conclusion}.

\section{Background}\label{sec:background}

We steadily developed, analyzed, and improved the \gls{sllbm} in recent years. In this section, we briefly recapitulate the principal development of the method.

Fluid dynamics can be described using either Eulerian or Lagrangian methods. Eulerian methods focus on describing the properties of the fluid at fixed points in space and time, while Lagrangian methods follow the motion of individual fluid particles over time.

As a variant of the Lagrangian description, semi-Lagrangian methods involve tracing the path of a fluid particle backward in time to determine its previous state and then using interpolation to obtain the fluid properties at that location to calculate the solution at the present time. When solving the Navier-Stokes equations, tracing the fluid particle back or forth leads to additional numerical tracking errors compared to standard schemes, thus influencing the quality of the solution \cite{Celledoni:2016}. By contrast, the characteristics of methods solving the Boltzmann equation are linear and, by design, derived methods avoid the tracking error.
One of these methods is the semi-Lagrangian \gls{lbm} (\acrshort{sllbm}), which has been developed by Kr\"amer \emph{et al.} \cite{Kraemer:2017} to constitute a fundamental extension of the regular \gls{lbm} convection and was applied and examined in later work by Kr\"amer \emph{et al.} \cite{Kramer2020}, di Ilio \emph{et al.} \cite{Ilio:2018} and Wilde \emph{et al.} \cite{Wilde2020,Wilde2021}.

The streaming step involves going back along the characteristics in time to determine the \acrlong{dp} and interpolating the distribution functions to obtain the corresponding departure distribution, replacing the \acrlong{ap} distribution. The collision step is performed using the interpolated distributions accordingly. One significant advantage of using this approach is the subsequent decoupling of time, space, and velocity discretization, allowing for much larger time-step sizes before instability occurs. Using the standard velocity space discretization (D2Q9, D3Q19, or D3Q27) with unity time step and regular non-stretched grid cells recovers the classic \gls{lbm}. Otherwise, the \acrlong{dp} is located off-grid and interpolation is required. 

Resolving compressibility effects requires stretching the discretization scheme beyond neighboring points. This typically results in huge velocity spaces and, hence, huge computational costs when simulating compressible flows in three dimensions. Therefore, many of the previous \gls{lbm} models for compressible flows found in the literature were limited to two dimensions. Wilde \emph{et al.} \cite{Wilde2021}, however, presented a vital extension of the method, emancipating it from simple two-dimensional flows. To reduce the number of discrete velocities, they exploited a major advantage of off-lattice Boltzmann methods, namely that the abscissae of the velocity sets can be freely arranged in space.
This allows the use of cubature rules to derive reduced velocity spaces and thus significantly improve performance, enabling the computation of the results presented in this paper.

\section{Compressible Semi-Lagrangian Lattice Boltzmann Method}\label{sec:method}

This work uses the compressible \acrfull{sllbm} proposed in a previous work by Wilde \emph{et al.} \cite{Wilde2020,Wilde2021a}. It is introduced in the following section.

\subsection{Flow Discretization}
\noindent The \gls{sllbm} equation reads
\begin{align}
        h_{i}(\mathbf x, t) = 
		& h_{i}(\mathbf x-\delta_{t} \boldsymbol \zeta_{i}, t-\delta_{t}) \\ 
  & - \frac{1}{\tau}  \left[h_{i}(\mathbf x-\delta_{t} \boldsymbol \zeta_{i}, t-\delta_{t})- h_{i}^{\mathrm{eq}}(\mathbf x-\delta_{t} \boldsymbol \zeta_{i}, t-\delta_{t}) \right] \nonumber \\ 
  & + \left(\frac{1}{\tau}-\frac{1}{\tau_\mathrm{Pr}}\right)h_i^*(\mathbf x-\delta_{t} \boldsymbol \zeta_{i}, t-\delta_{t}) \nonumber
\end{align}
where $\delta_t$ is the time-step size and $h_i\in \{f_i,g_i\}$ represents two sets of discrete populations with the equilibrium counterparts $h_i^\mathrm{eq}\in \{f_i^\mathrm{eq},g_i^\mathrm{eq}\}$, respectively, where $i \in [0,Q-1]$, with $Q$ the number of distribution functions (see Sec.~\ref{sec:cubature}). The propagation of the distribution functions occurs along the characteristic directions determined by the discrete velocities $\boldsymbol \zeta_i$, see Sec.~\ref{sec:cubature}. Consequently, the \gls{sllbm} starts at an \acrfull{ap} $\mathbf x_k = \mathbf x_{AP}$ and follows the trajectory backward in time to determine the \acrfull{dp} $\mathbf x_k-\delta_t \boldsymbol \zeta_i = \mathbf x_{DP}$.

The relaxation parameter $\tau=\mu / p$ is the ratio of dynamic viscosity $\mu$ and local pressure $p=\rho R T$, with density $\rho$ and temperature $T$, whereas the second relaxation parameter $\tau_\mathrm{Pr}=(\tau-0.5)/\mathrm{Pr}+0.5$ is dependent on the Prandtl number $\mathrm{Pr}$. To allow for non-unity Prandtl numbers, quasi-equilibria $h_i^*$ are calculated following \cite{Ansumali:2007} and \cite{Wilde2021a}. The conserved moments of density, momentum, and total energy read

\begin{align}
    \rho&=\sum_{i=0}^{Q-1} f_i \label{eq:density} , \\
    \rho \mathbf{u} &= \sum_{i=0}^{Q-1} \boldsymbol{\zeta}_{i} f_i \label{eq:momentum} ,\mathrm{ and} \\
     2\rho c_v T &= \sum_{i=0}^{Q-1} \left(|\boldsymbol{\zeta}_i|^2 f_i  \!+ g_i \right) .
\end{align} 
The discrete equilibrium values $f_i^{eq}$ are determined through
\begin{equation}
    f_i^{\mathrm{eq}}(\mathbf{x},t) = w_i \sum_{n=0}^N \frac{1}{n!}\boldsymbol{\alpha}^{(n)}_{eq}(\mathbf{x},t) :  \boldsymbol{\mathcal{H}}_i ^{(n)},
\end{equation}
where the expansion order is set to $N=4$ to enable compressible simulations \cite{shan2006kinetic}. The expansion coefficients $\boldsymbol{\alpha}^{(n)}_{eq}$ and the Hermite coefficients $\boldsymbol{\mathcal{H}}_i ^{(n)}$ are listed in \cite{Wilde2020} and omitted here for brevity. For the context of Hermite coefficients, refer to Sec.~\ref{sec:cubature}. The equilibrium $g_i^\mathrm{eq}$ is determined by
\begin{equation}
    g^\mathrm{eq}_i = T (2 c_v - D) f^\mathrm{eq}_i~,
\end{equation}
depending on the specific heat capacity at constant volume, $c_v$, and dimension $D$.

The computational domain is discretized using $M_\Upsilon$ cells, with $\Upsilon$ indicating each cell, which includes $N_\Upsilon$ preset discretization points $\mathbf x_j$, $j\in[1,N_\Upsilon]$. A set of $Q$ distribution functions $\hat h_{i\Upsilon j}$, $i \in [0,Q-1]$, is associated with each $\mathbf x_j$. The off-vertex distribution function values $h_i(\mathbf x,t)$ are reconstructed using polynomial interpolation in each cell. A polynomial basis is defined by making use of Lagrange polynomials $L_{\Upsilon j}$ of order $p$ and distribution functions $\hat h_{i\Upsilon j}$ defined at the support points of each cell $\Upsilon$, such that
\begin{equation}\label{eq:intf}
    h_i(\mathbf{x},t) = \sum_{j=1}^{N_\Upsilon}\hat{h}_{i\Upsilon j}(t) L_{\Upsilon j}(\mathbf{x})
\end{equation}
for every position $\mathbf x$ within $\Upsilon$. This is then used to calculate the distribution functions at the \acrlong{ap} at $\mathbf x_j = \mathbf x_{AP}$, with a \acrlong{dp} located in a cell $\Upsilon'$.

\begin{equation}
    h_{i\Upsilon j}(\mathbf{x}_{AP},t) = 
    h_{i}(\mathbf{x}_{AP}-\delta_t \boldsymbol\zeta_i,t-\delta_t) = \sum_{j=1}^{N_\Upsilon} \hat{h}_{i\Upsilon' j}(t-\delta_t) L_{\Upsilon' j}(\mathbf{x}_{AP}-\delta_t \boldsymbol\zeta_i)
\end{equation}

\subsection{Velocity space discretization}\label{sec:cubature}

The \gls{lbm} velocities $\boldsymbol \zeta_i$ combined define the velocity space discretization or stencils. They are referred to according to dimensionality $D$ and number of abscissas $Q$. Incompressible \gls{lbm}, for example, typically uses D3Q19 and D3Q27 stencils. Stretching the velocity space requires a higher-order quadrature and, therefore, very large stencils, previously limiting the method to two dimensions. One well-known exception is the work by Frapolli \emph{et al.} \cite{Frapolli:2016}, who used a D3Q343 discretization of the velocity space based on a D1Q7 using product rules.

The \textit{order of error} depends on the quadrature degree $\mathfrak{N}$. Incompressible \gls{lbm} requires $\mathfrak N \ge 5$, while compressible \gls{lbm} requires $\mathfrak N \ge 8$. On $n$ equidistant points, Newton-Cotes quadrature achieves the quadrature degree $\mathfrak{N}\le n-1$, while Gauss-Hermite quadrature uses the roots of the $n$-th Hermite polynomial $\mathcal{H}^{(n)}$ with non-equidistant points for quadrature degree $\mathfrak{N}\le 2n-1$. This reduces the velocity space from $Q=7$ to $Q=5$ in 1D for $\mathfrak{N}=9$. Using product rules, this results in a 60\% decrease in 3D, i.e. from $Q=343$ to $Q=125$.

For the multivariate quadrature in 3D, cubature rules can be used instead of product rules to further reduce the velocity space. Stroud \cite{stroud1960quadrature} showed the theoretically minimal required number of quadrature points to be
\begin{equation}
    Q\le
    \binom{\lfloor\mathfrak N / 2 + D \rfloor}{
        \lfloor \mathfrak N / 2 \rfloor},
\end{equation}
with $\lfloor x \rfloor$ the floor function of $x$ and ${\tiny \binom{a}{b}}$ the binomial coefficient. While this gives a theoretical limit of 15 and 35 for $\mathfrak{N}=9$, until now, cubature formulas were found with a minimum number $Q$ of 19 in 2D and 45 in 3D \cite{cools2003encyclopedia}.
Consequentially, Wilde \emph{et al.} \cite{Wilde2021} were able to apply cubature rules to derive (with $\mathfrak{N}=9$) D2Q19 and D3Q45 stencils, which were used throughout this work, reducing the velocity space by up to 86\%.

\subsection{Boundary conditions}\label{sec:BC}

Near the edges of the simulation domain, the \acrlong{dp} may lie outside the domain. In this case, the boundary condition applied will depend on the type of boundary that the trajectory has hit. Three different boundary condition types have been developed for this work: equilibrium boundary conditions for inlets, bounce-back boundary conditions for solid walls, and zero-gradient boundary conditions for outlets.

\subsubsection{Equilibrium boundary conditions}\label{sec:eqbc}

Equilibrium boundary conditions are sufficient for completely supersonic inlets because the flow information, including sound waves, can only travel downstream, and the flow is assumed to be in equilibrium. Each subsection of Sec.~\ref{sec:results} shows a use case of this boundary condition type.

Using the density $\rho_w$, velocity $\boldsymbol{u}_w$, and temperature $T_w$ at the wall, the equilibrium distribution function value $f_i^\mathrm{eq}$ can be determined. This equilibrium distribution function value $f_i^\mathrm{eq}$ is directly assigned to the corresponding \acrlong{ap}.

\begin{equation}
    f_i(\boldsymbol{x}_\mathrm{AP},t) = f_i^\mathrm{eq}(\rho_w, \boldsymbol{u}_w, T_w)
\end{equation} 

\subsubsection{Bounce-back and isothermal boundary conditions}\label{sec:bbbc}

No-slip boundary conditions can be modeled using bounce-back boundary conditions. For the \gls{sllbm}, a combination of the standard \gls{lbm} half-way bounce-back \cite{Frisch1986,Ginzbourg1994} and interpolation-based bounce-back boundary conditions \cite{Bouzidi2001} is used in Sec.s~\ref{sec:2dnaca} to \ref{sec:channel}. Once a trajectory hits a boundary condition, the remaining distance will be traveled back in the opposite direction until the \acrlong{ap} at $\boldsymbol{x}_\mathrm{AP}$ is reached. The opposite distribution function value of the \acrlong{dp}, $f_{\Bar{i}}(\boldsymbol{x}_\mathrm{DP})$, is assigned to the distribution function value at the \acrlong{ap}.
\begin{equation}
    f_i(\boldsymbol{x}_\mathrm{AP},t) = f_{\Bar{i}}(\boldsymbol{x}_\mathrm{DP},t-\delta_t),
\end{equation}
with 
\begin{equation}
    \boldsymbol{x}_\mathrm{DP} =  \boldsymbol{x}_\mathrm{W} + \boldsymbol\iota , \qquad \boldsymbol\iota=\boldsymbol{\zeta}_i \delta_t - (\boldsymbol{x}_\mathrm{AP} - \boldsymbol{x}_\mathrm{W} ) .
\end{equation}
Many compressible flow problems require additional temperature boundary conditions. One example is the turbulent supersonic channel flow simulation presented in Sec.~\ref{sec:channel}. Isothermal boundary conditions are required to keep the walls cooled to a temperature $T_w$, to allow for supersonic flow. This can be achieved by proposing a novel modification of the distribution functions after one of the trajectories has collided with the wall. In our realization, this modification affects all $Q$ distribution function values at each grid point near the wall, and the temperature at the \acrlong{ap} is set to the wall temperature $T_w$.
The \acrlong{ap} values are used for the density $\rho$ and the velocities $\boldsymbol{u}$, then the equilibrium distributions are used to set 
\begin{equation}\label{eq:thermalBC}
\vspace{-.2cm}
    f_i(\boldsymbol{x}_\mathrm{AP},t) = f_{\Bar{i}}(\boldsymbol{x}_\mathrm{DP},t-\delta_t) +  f^\mathrm{eq}_i(\rho,\boldsymbol{u},T_w) - f^\mathrm{eq}_i\left(\rho,\boldsymbol{u},T(\boldsymbol{x}_\mathrm{AP})\right).
\end{equation}
This modification preserves the conservation of mass and momentum. Additionally, it retains all shear moments compared to equilibrium boundary conditions if the temperature boundary condition is combined with the bounce-back boundary condition above.

\subsubsection{Outlet boundary conditions}\label{sec:outletbc}

One way to realize outlet boundary conditions is to set the gradient of the distribution functions to zero. If an outlet boundary condition is detected when determining the \acrlong{dp}, the search for the \acrlong{dp} is restarted from the \acrlong{ap} in the opposite direction until the distance $\boldsymbol{\zeta}_i \delta_t$ has been covered. I.e., with $\boldsymbol{\zeta}_i$ the discrete velocity vector of the $i$-th distribution function and $\delta_t$ the time step,
\begin{equation}
    f_i(\boldsymbol{x}_\mathrm{AP},t) = f_i(\boldsymbol{x}_\mathrm{AP}+\boldsymbol{\zeta}_i \delta_t,t-\delta_t).
\end{equation}
This boundary condition is demonstrated in Sec.~\ref{sec:2dnaca} to \ref{sec:channel}.

\section{Results}\label{sec:results}

\noindent Several test cases were evaluated using the \gls{sllbm} approach to investigate its applicability and suitability to efficiently deal with various compressibility effects and turbulence. They are summed up in this section. The simulations were done on the OMNI cluster at the University of Siegen on nodes consisting of two AMD EPYC 7452 CPUs with 32 cores, 128 MB cache, and 256 GB DDR4 RAM per node.

\subsection{Supersonic 2D NACA-0012 profile}\label{sec:2dnaca}

The first investigated scenario is the supersonic flow around a NACA-0012 airfoil with Mach number $\mathrm{Ma}=1.5$ at an angle of attack of $\mathrm{AoA}=0^\circ$. The Reynolds number has been set to $\mathrm{Re}=\rho_0 U C/\mu = 10{,}000$ based on the free stream velocity $U$ and the chord length $C$ of the airfoil. The configuration was consistent with the work of Hafez and Wahba \cite{Hafez2007}, Latt \emph{et al.} \cite{Latt2020}, and Frapolli \emph{et al.} \cite{Frapolli2020} and chosen to facilitate comparisons with those investigations. Tab. \ref{tab:NACA} summarizes the used parameters.

\begin{table}[b!]
    \centering
    \caption{Parameters of the 2D NACA-0012 test case. \vspace{-.15cm}}
    {\small \begin{tabular}{ll|l}
    Parameter & Abbr. & value \\
    \hline
    Mach number     & $\mathrm{Ma}$ & $1.5$         \\
    Reynolds number & $\mathrm{Re}$ & $10{,}000$  \\
    Order of finite element                &   $p$ & 4            \\
    Number of grid points & $N$ & $1.86$ million      \\
    time-step size & $\delta_t$ & $0.00015$ \\
    Number of time steps & $N_t$ & $200{,}000$ \\
    Angle of Attack & $\mathrm{AoA}$ & $0^{\circ}$ \\
    \end{tabular}}
    \label{tab:NACA}
\end{table}

\begin{figure}[!t]
    \centering
    \includegraphics[width=0.45\linewidth]{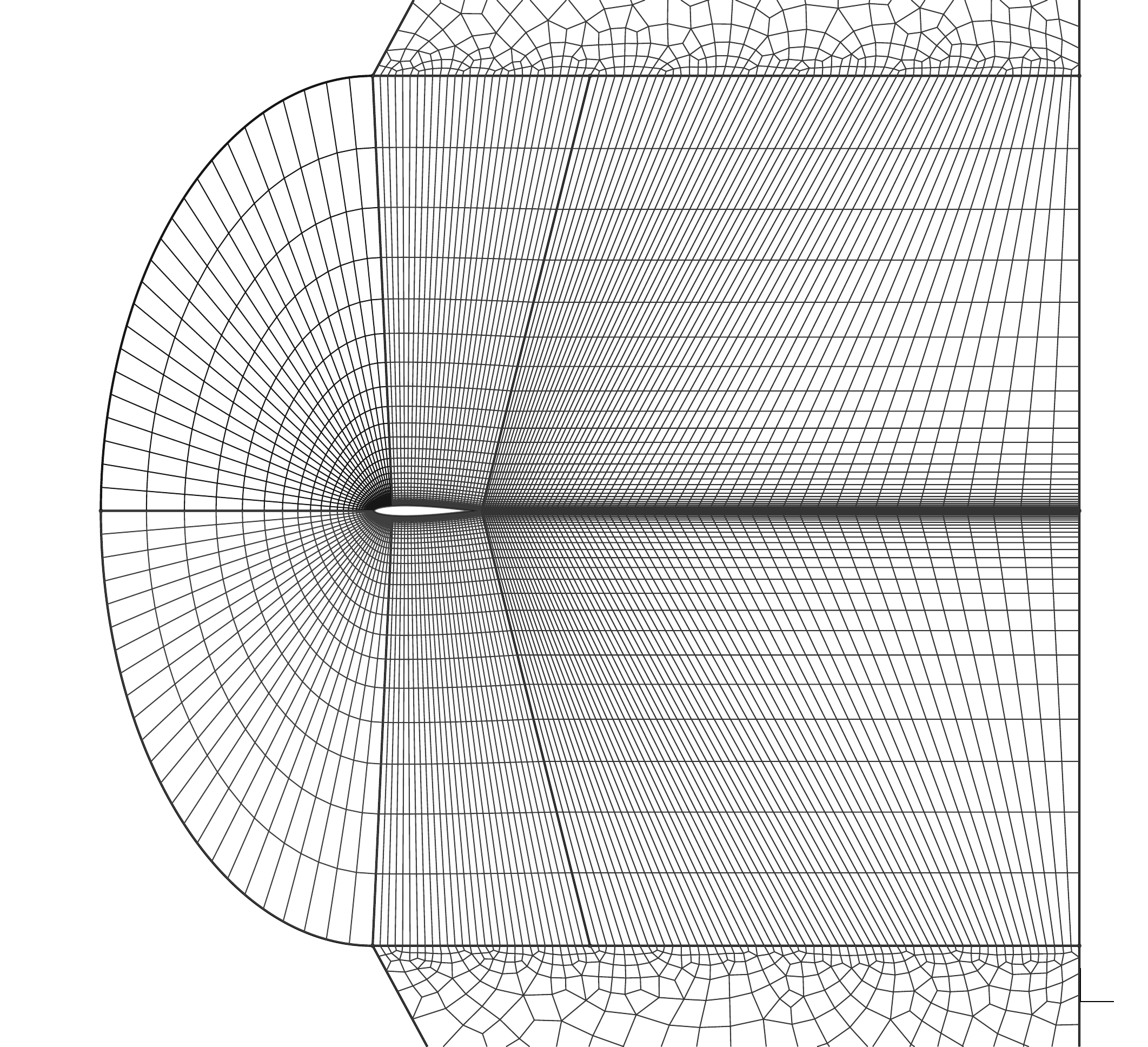}
    \caption{Principal mesh design for the supersonic flow around the 2D NACA-0012 airfoil. The mesh was further refined for the actual simulation (taken from \cite{wilde2023}).}
    \label{fig:mesh}
\end{figure}

Fig.~\ref{fig:mesh} shows the inner part of the structured, body-fitted mesh. It was generated using the software package \textit{Gmsh} \cite{Geuzaine2009} and loaded into the finite element package \textit{Deal.ii} \cite{Heltai:2021}, which constitutes the backbone of the \gls{sllbm} solver \cite{Kramer2020}. The mesh was further refined for the present simulation and the order of the finite elements was set to $p=4$, yielding 1.86 million grid points in total. The time-step size was $\delta_t=0.00015$ and the simulation ran 200{,}000 time steps until $t=30t^*$, with $t^*$ the characteristic time $t^*=C/U$. With the specified configuration, the simulation required three nodes for six hours of computing time.

The equilibrium boundary condition with prescribed pressure and velocity (described in Sec.~\ref{sec:eqbc}) was used on the inlet plane. On the outlet plane, the outlet boundary condition described in Sec.~\ref{sec:outletbc} was applied. Additionally, both domain boundaries in the lateral directions normal to the wing chord were extended using a sponge zone, consisting of unstructured grid cells with increasing grid size to dampen waves and to guide all shocks out of the simulation domain in such a way that no spurious shock reflections could occur. The velocity space was discretized using the D2Q19 velocity set with $\mathfrak{N}=9$. At the beginning of the simulation, the flow domain was globally superimposed with a sine wave perturbation $u_y = 0.1\cdot u_x\sin(1.3y+0.1)$ to excite a vortex street in the wake.

To quantify the results, the pressure coefficient was measured in front of and behind the airfoil as well as on its surface. The pressure coefficient $C_P$ for compressible flows is defined as
\begin{equation}
    C_P=\frac{p-p_0}{\frac{\gamma}{2}p_0\mathrm{Ma}^2},
\end{equation}
where $p_0$ is the reference pressure prescribed at the inlet.
 \begin{figure}[b!]
    \centering
    \includegraphics[width=0.7\textwidth]{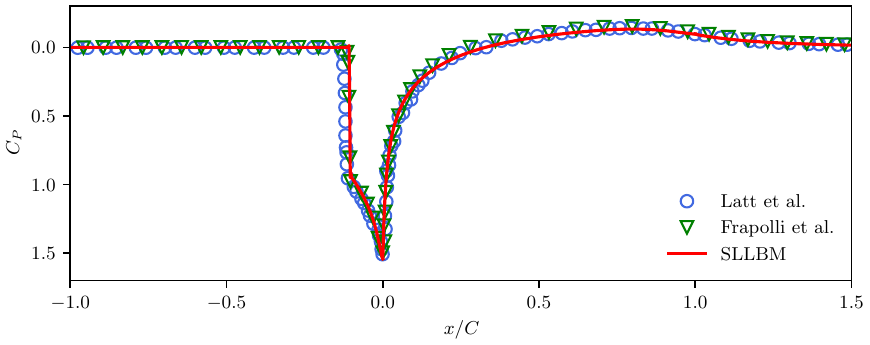}
    \vspace{-.2cm}
    \caption{Pressure coefficient $C_P$ along the airfoil cord compared to \cite{Latt2020} and \cite{Frapolli2020}. $x/C$ is positive downstream.}
    \label{fig:naca_cp}
\end{figure}
Fig.~\ref{fig:naca_cp} shows the good agreement of the pressure coefficient along the chord in comparison to results of simulations from Latt \emph{et al.}\cite{Latt2020} and Frapolli \emph{et al.}\cite{Frapolli2020}.

\begin{figure}[ht!]
    \centering
    \begin{subfigure}[t]{0.49\textwidth}
        \includegraphics[width=\linewidth]{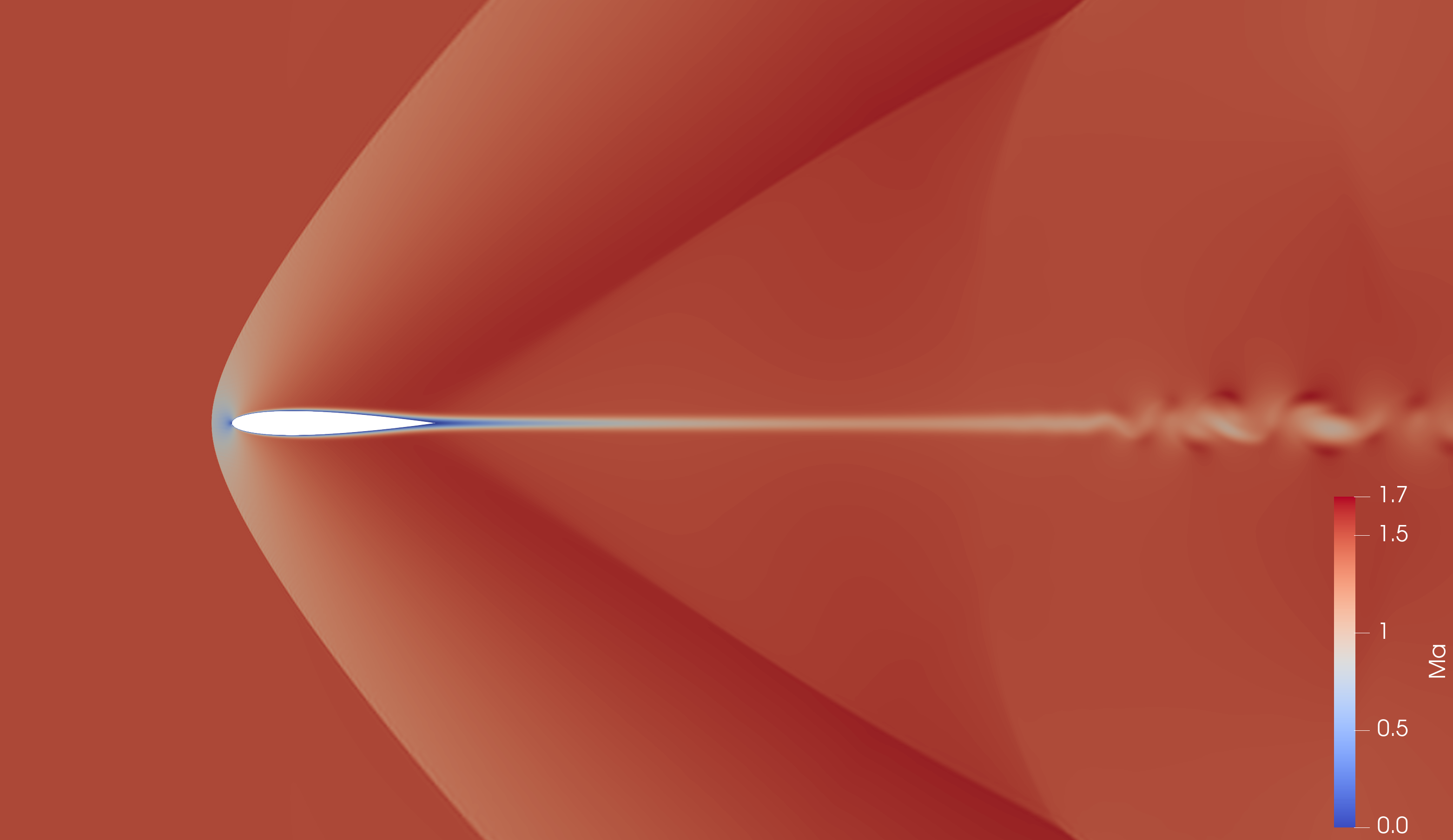}
        \caption{Vortex street in the velocity profile of the transient state at $\tau=7.92$, normalized to the speed of sound.}
        \label{fig:naca_Ma}
    \end{subfigure}
    \begin{subfigure}[t]{0.49\textwidth}
        \includegraphics[width=\linewidth]{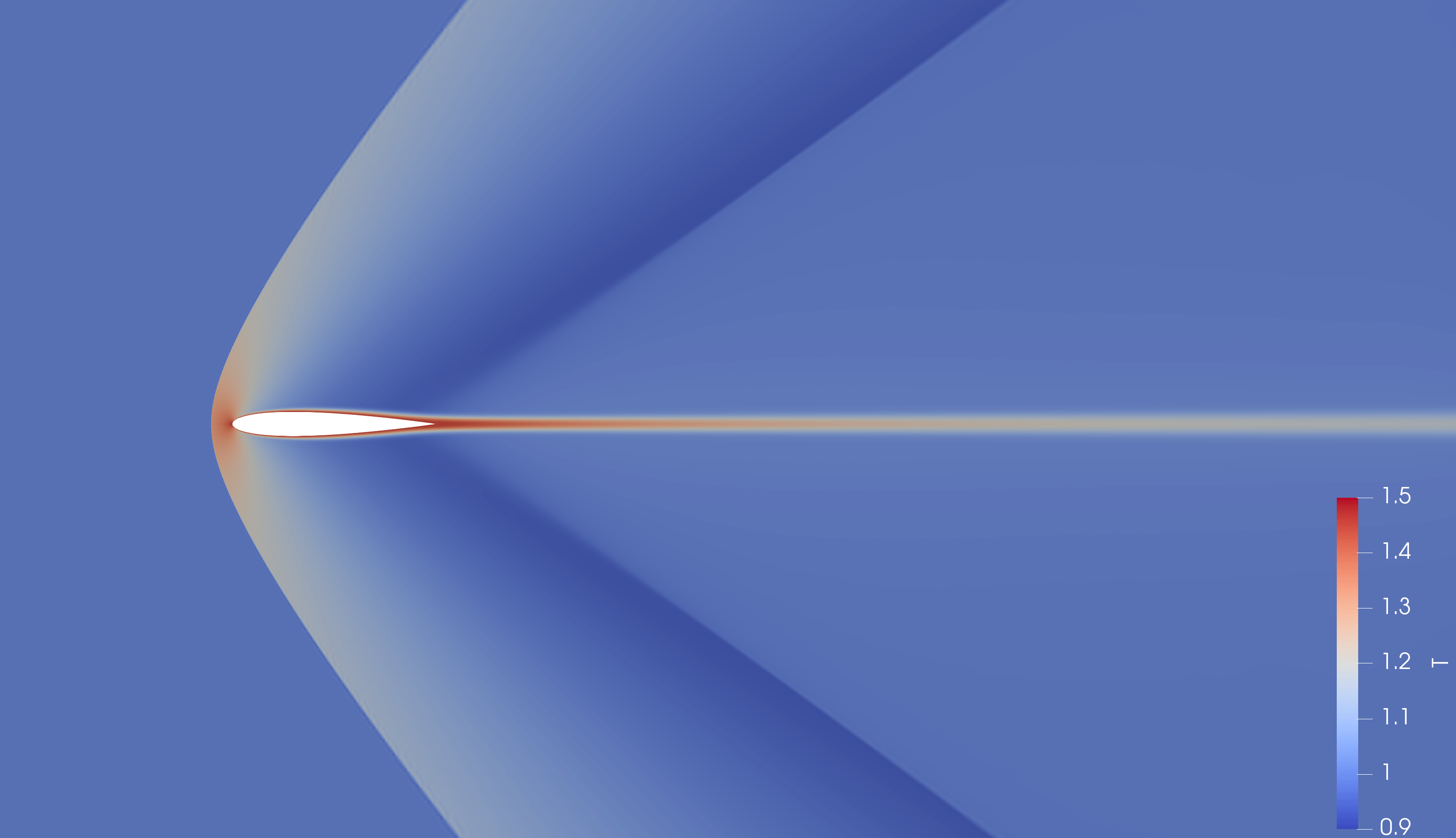}
        \caption{Temperature profile without vortex street in the converged state at $\tau=19.80$.\\\phantom{ }}
        \label{fig:naca_T}
    \end{subfigure}
    \caption[Supersonic 2D flow profiles around a NACA-0012 aifoil]{Supersonic 2D flow profiles around a NACA-0012 aifoil, comparing the transient state at $\tau=7.92$ to the converged state at $\tau=19.80$, and velocity and temperature. The vortex street (\textit{left}) is only present before the converged state (\textit{right}).}
\end{figure}

Fig.~\ref{fig:naca_T} shows the temperature profile of the converged flow. The curved shock in front of the airfoil, which is spatially separated from the body, is clearly visible. Starting from the trailing edge, another shock is formed with a more acute angle with respect to the leading edge. Overall, the visual agreement with the results of Latt \emph{et al.} \cite{Latt2020}, Hafez and Wahba \cite{Hafez2007}, and Frapolli \emph{et al.} \cite{Frapolli2020} is very good. 

Towards the outlet, however, the results differ from Frapolli \emph{et al.} and Latt \emph{et al.}, who observed a vortex street in the flow near the outlet in their final state. While we obtain a vortex street in the transient phase, as shown in  Fig.~\ref{fig:naca_Ma} at time $t=6.5$, the converged wake structure does not contain a stable vortex street. This is consistent with the observations of Hafez and Wahba \cite{Hafez2007}.

To investigate the cause of this deviation, further configurations with angles of attack $\mathrm{AoA}\in \{1^{\circ},2^{\circ},3^{\circ},4^{\circ}\}$ as well as $\mathrm{Re}=20{,}000$ were calculated. Here, too, a stable vortex street without further disturbances did not form.
The same is true when increasing the order of the approach functions to $p=6$ or using the D2Q25 velocity space with $\mathfrak{N}=9$. The vortex street could only be observed again when perturbations were present in the simulation domain, such as an insufficiently resolved shock or poor meshing. 
Therefore, two explanations are possible for the deviation from the work of Frapolli \emph{et al.} and Latt \emph{et al.}:
\begin{enumerate}
    \item The body-fitted grid of the \gls{sllbm} simulations excellently resolves the fine trailing edge of the airfoil, while the aforementioned authors calculated with a Cartesian stepped grid. This might lead to disturbances triggering the vortex street.
    \item The \gls{sllbm} simulation did develop a vortex street but was continued further until convergence. This convergence may have been achieved in longer simulations, even in a stepped grid.
\end{enumerate}

\subsection{Supersonic 3D Sphere}\label{sec:3dsphere}

\begin{figure}[t!]
    \centering 
    \vspace{-.1cm}
    \includegraphics[width=0.35\linewidth]{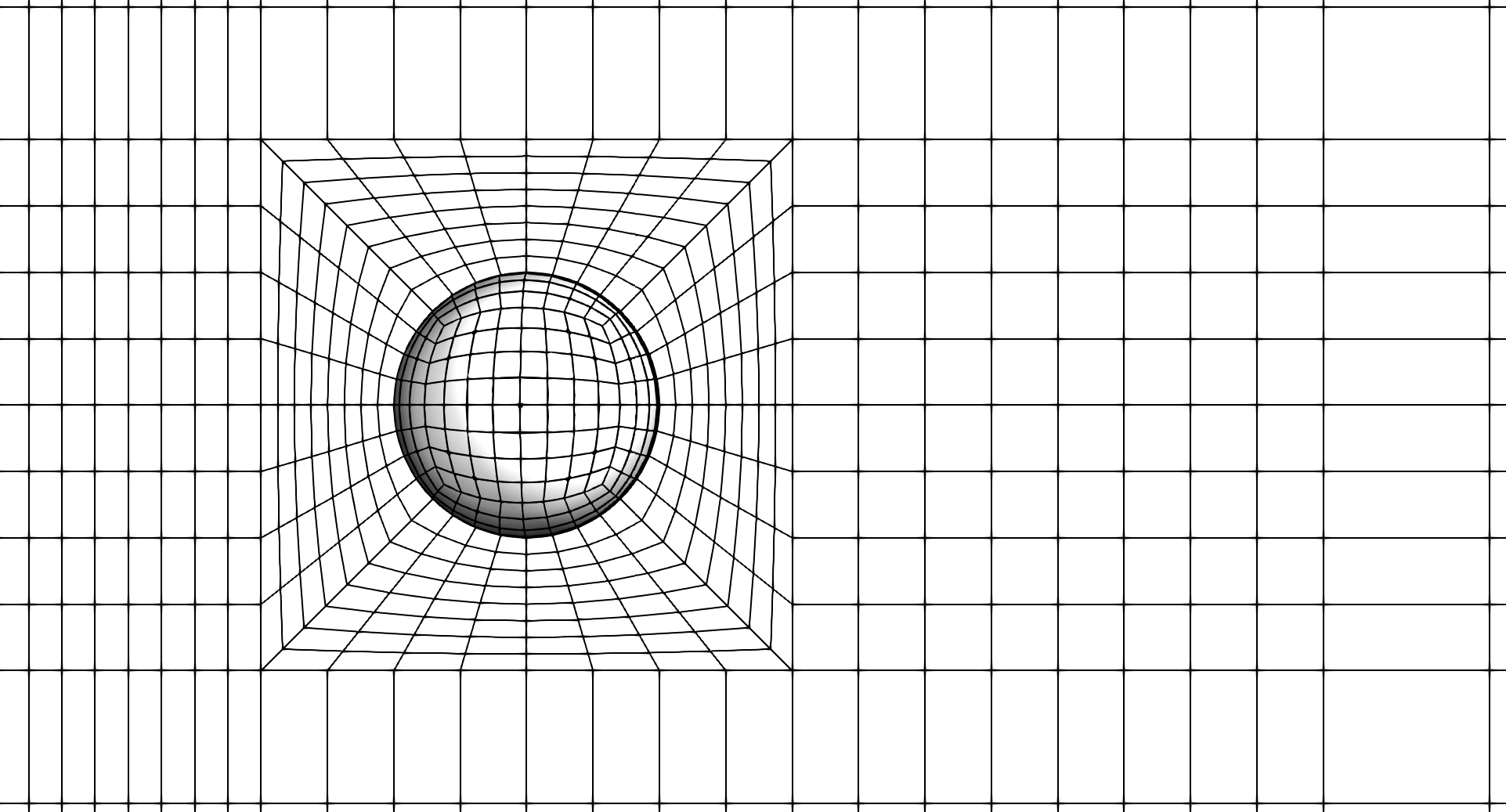}\vspace{-0.1cm}
    \caption{Vizualisation of a selected section of the numerical mesh for the supersonic flow around a solid sphere (taken from \cite{wilde2023}). Shown is the x-y plane at $z=0$ and the mesh along the sphere's surface. The mesh was further refined for the actual simulations. }
    \label{fig:sphere_grid}
\end{figure}

To show the capability of this approach to simulate 3D supersonic flows, this section presents the supersonic flow around a solid sphere. For the simulation, a body-fitted mesh was generated along a sphere and embedded into the surrounding Cartesian grid \cite{Heltai:2021} using \textit{Deal.ii}. The sphere center is located at the coordinate origin and the domain has a spatial dimension, normalized by sphere diameter $D$ of $x\in[-2,8],(y,z)\in[-5,5]$. The circumference of the sphere is resolved with 128 grid points, while the first grid point of the fluid in the radial direction is located at $r^+=0.01$, with respect to the surface of the sphere. A section of the mesh is shown in Fig.~\ref{fig:sphere_grid}.

Equilibrium boundary conditions with given density $\rho=1$, temperature $T=1$, and the respective Mach number $\mathrm{Ma}$ of the simulation were used for all boundaries except the outlet, where outlet boundary conditions (Sec.~\ref{sec:outletbc}) were applied. The ratio of the specific heats was set to $\gamma = 1.4$ and the Prandtl number to $\mathrm{Pr}=0.71$. The Reynolds number was fixed at $\mathrm{Re}=300$, meaning that the flow can just be assigned to the continuum region at a Mach number of $\mathrm{Ma}=1.5$, following the definition of the Knudsen number $\mathrm{Kn}=\sqrt{\gamma \pi /2}\mathrm{Ma}/\mathrm{Re}=0.007$. 
The velocity space was discretized using the D3Q45 velocity space. The order of the finite elements was chosen as $p=4$. A total of 1.38 million grid points were used with a time-step size of $\delta_t=0.0084$. With the specified configuration, the simulation up to $t=25$ (equalling 8100 time steps) required about seven hours of computing time on three nodes.

\begin{figure}[b!]
    \centering
    \vspace{-.2cm}
    \includegraphics[width=0.73\linewidth]{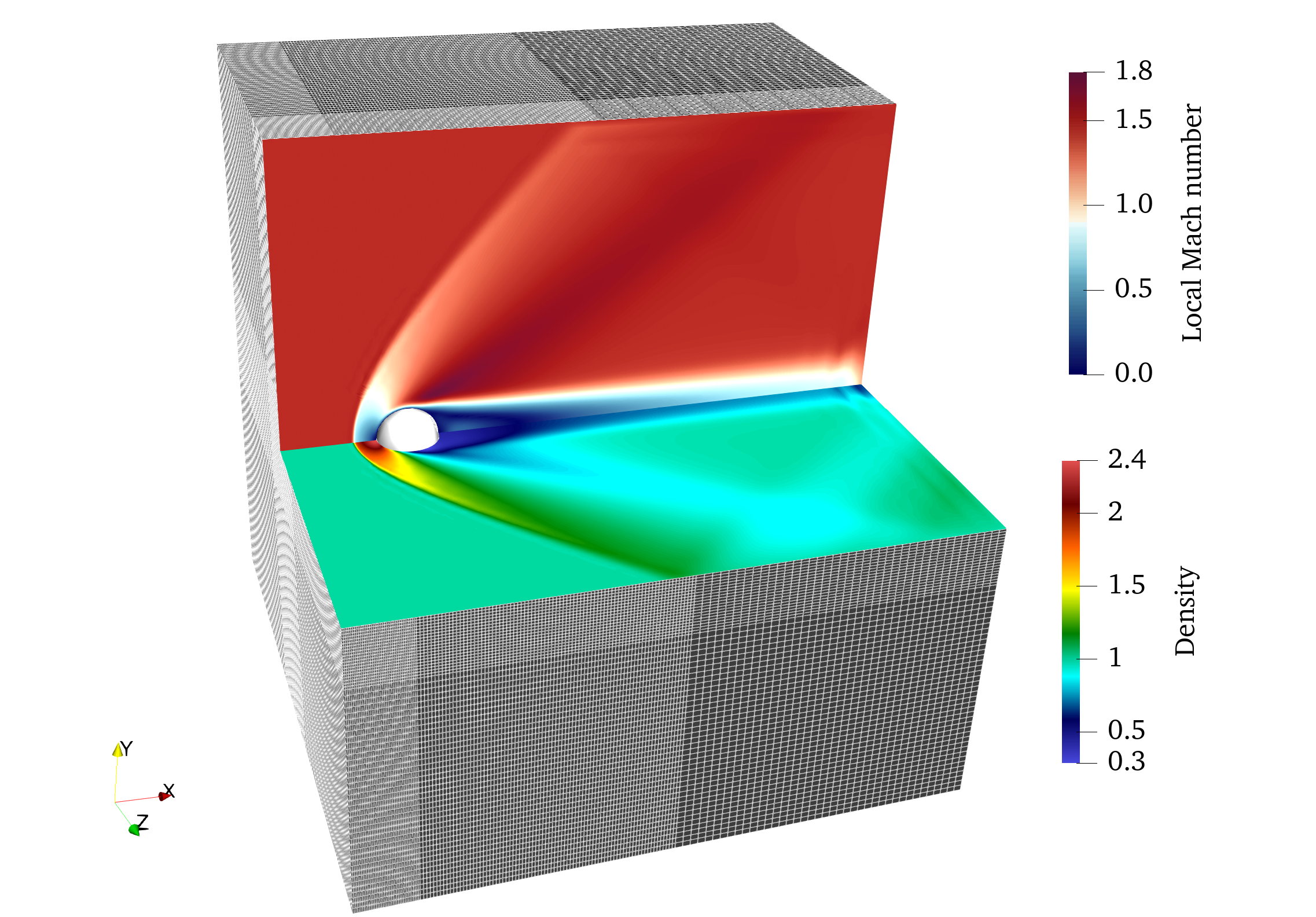}
    \caption{Mach number (x-y plane) and temperature (x-z plane) of the flow around a sphere at Mach number $\mathrm{Ma}=1.5$. Below the sectional planes, the grid is shown}
    \label{fig:sphere}
\end{figure}

Fig.~\ref{fig:sphere} shows the Mach number and temperature at $t=30$. The settled shock and the heated wake of the flow can be clearly identified. Due to the low Reynolds number, the wake is stable, as also reported by Nagata \emph{et al.} using a radially resolved grid of similar size (0.91 Mio. grid points) with a \gls{weno} method \cite{Nagata:2016}.
Wave reflections are visible close to the outlet, which can be explained by the rather simple outlet boundary condition. They did not affect the presented results at this stage. Using characteristic boundary conditions \cite{heubes2014characteristic} or adding a strongly stretched buffer region \cite{Nagata:2020} would likely mitigate this effect.

\begin{figure}[t!]
    \centering
    \includegraphics[width=0.6\textwidth]{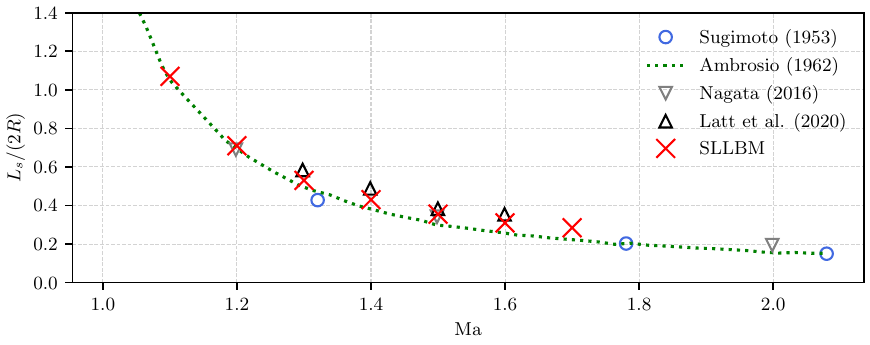}
    \vspace{-.2cm}
    \caption{Distances of the shock front from the sphere surface with the \gls{sllbm} at $\mathrm{Re}=300$, compared to experimental and numerical data in Hida \cite{Hida:1953}, Ambrosio \& Wortman \cite{Ambrosio:1962}, Nagata \emph{et al.} \cite{Nagata:2016} and Latt \& Coreixas \cite{Latt2020}.}
    \label{fig:distance}
\end{figure}

To validate the results, the distance of the shock front from the sphere surface through the symmetry plane as a function of the Mach number was measured. Overall, it is in very good agreement with the available literature, as shown in Fig.~\ref{fig:distance}. Additionally, it seems to be closer to the literature than the \gls{lbm} results of Latt \emph{et al.}  \cite{Latt2020}, who approximated the sphere without body-fitted grids and required a much larger resolution of 8 million grid points. 

Simulations with inlet Mach numbers $\mathrm{Ma}\geq1.8$ became unstable using the \gls{sllbm} in the given configuration, showing the limitations of the present approach. This behavior can be explained by local Mach numbers aproaching $\mathrm{Ma}\approx 2.0$ due to the fluid acceleration when flowing around the sphere body.

\begin{figure}[ht!]
\vspace{-.3cm}
    \centering
    \includegraphics[width=0.67\linewidth]{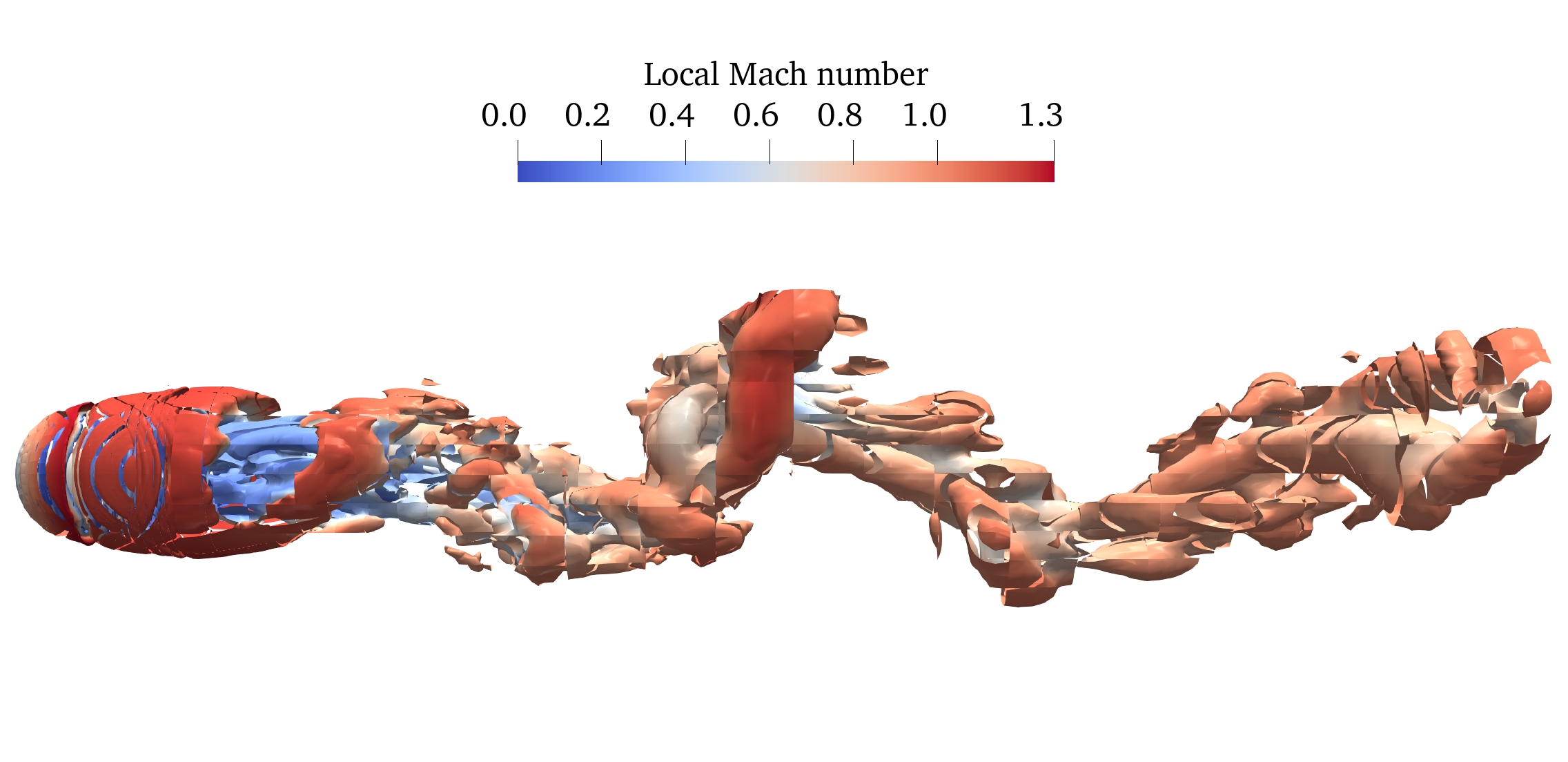}
    \vspace{-.99cm}
    \caption{3D vortex structure of the transonic flow around a sphere at $\mathrm{Ma}=1.05$ and $\mathrm{Re}=1000$ after $t=24.6$. Shown is the isocontour of the Q criterion at $\hat{Q}=0.0005$, colored with Mach number.}
    \label{fig:Sphere-1000}
\end{figure}

\begin{figure}[b!]
    \centering
    \includegraphics[width=0.7\linewidth]{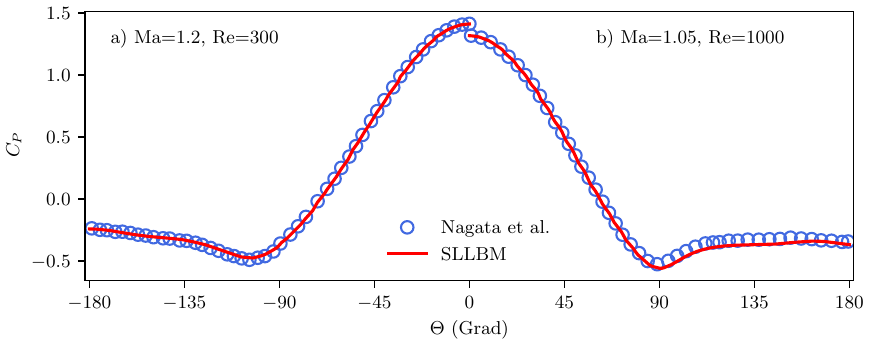}
    \vspace{-.3cm}
    \caption{Progression of pressure coefficient $C_P$ along the sphere surface for $\mathrm{Ma}=1.2, \mathrm{Re}=300$ (left) and $\mathrm{Ma}=1.05, \mathrm{Re}=1000$ (right). Reference from Nagata \emph{et al.} \cite{Nagata:2020}}
    \label{fig:sphere_pressure}
\end{figure}

In additional simulations, the Reynolds number was increased to $\mathrm{Re}=1000$ to compare to other literature results. It was observed in the past that the vortex street disappears rapidly for transonic and supersonic flows (cf. \cite{Nagata:2020}). Therefore, the Mach number was set to $\mathrm{Ma}=1.05$ to place the configuration into the transonic regime and match the reference case. A new domain size of  $x\in[-4,12],(y,z)\in[-8,8]$ was chosen, to facilitate the full wake structure, but by keeping the spherical surface resolution constant, resulting in an increased number of grid points of 4.34 million and a decreased time-step size of $\delta_t=0.003$. Fig.~\ref{fig:Sphere-1000} depicts the isosurface using the Q criterion at $\hat{Q}=0.0005$ and time $t=24.6$. The vortex street with relatively long vortex tubes is nicely illustrated and qualitatively corresponds well to the corresponding figure in Nagata \emph{et al.} \cite{Nagata:2020}.
The pressure coefficient $C_P$ along the sphere surface is shown exemplarily in Fig.~\ref{fig:sphere_pressure}, for the cases $\mathrm{Ma}=1.2, \mathrm{Re}=300$ and $\mathrm{Ma}=1.05, \mathrm{Re}=1000$, compared to the literature \cite{Nagata:2020}. Very good agreement with the reference cases over the entire spherical surface is obtained.

\subsection{Supersonic 3D Channel Flow at $\mathrm{Ma}=1.5$} \label{sec:channel}

A three-dimensional fully developed compressible supersonic channel flow was simulated for the first time in \gls{lbm} literature, to the best of our knowledge. This configuration constitutes an important canonical test case to investigate variable property compressibility effects \cite{Foysi:2004} and allows for detailed comparisons of various statistics with different numerical solvers. 

In Sec.~\ref{sec:channel} and \ref{sec:mixing}, Reynolds and Favre averages are used. Reynolds averages are defined as the arithmetic mean across the self-similar dimensions $x$ and $z$ and denoted as
\begin{equation}\label{eq:reynolds}
    \phi = \overline{\phi} + \phi'
\end{equation}
with $\phi$ a variable, $\overline{\phi}$ the Reynolds average, and $\phi'$ the Reynolds fluctuations. Taking into account the larger compressibility effects of high Mach number cases, mass-weighted Favre averages \cite{favre1983turbulence} are typically used for compressible flow \cite{Foysi:2004}. They are similarly defined as
\begin{equation}\label{eq:favre}
    \phi = \tilde{\phi} + \phi'', \qquad \tilde{\phi}=\frac{\overline{\rho\phi}}{\overline{\rho}}, \qquad \overline{\rho\phi''}=0.
\end{equation}

The bulk Reynolds number and Mach number of the test case were chosen to coincide with those of comparable cases in the literature. A global bulk Mach number of $\mathrm{Ma}_b = u_b/c_w = 1.5$ was investigated, with $c_w$ being the speed of sound at the wall and $u_b$ the bulk velocity, defined as
\begin{equation}\label{channel:bulk}
    u_b = \frac{1}{\delta_c}\int_0^{\delta_c} \overline{u} dy,
\end{equation}
making use of the channel half width $\delta_c$.
Therefore, the Reynolds was set as
\begin{equation}
    \mathrm{Re}_b = \frac{\rho_b u_b \delta_c}{\mu_w} = 3000, 
\end{equation}
using the bulk density\\[-.5cm] 
\begin{equation}\label{channel:momentum}
    \rho_b = \frac{1}{\delta_c}\int_0^{\delta_c} \overline{\rho} dy 
\end{equation}
and the dynamic viscosity at the wall $\mu_w$. In the following, the subscript $w$ always denotes the values at the walls of the channel. The friction Reynolds number $\mathrm{Re}_\tau$, defined as
 $\mathrm{Re}_\tau = {\rho_w u_\tau \delta_c }/{ \mu_w}$,
was determined to be $\mathrm{Re}_\tau = 220.6$, which agrees well with values typically in the literature when using the above global parameters. The dynamic viscosity follows a power law of the form 
\begin{equation}
  \mu = \mu_w \left(\frac{T}{T_w}\right)^{0.7}. 
\end{equation}

The isentropic exponent is set as $\gamma=1.4$ and the Prandtl number as $\mathrm{Pr}=0.7$, making the configuration equivalent to comparable research \cite{Foysi:2004, Coleman:1995, Gosh:2010, Ruby:2022}.

The flow domain normalized with $\delta_c$ measured $4\pi\times 2\pi\times 2\pi$, in $x$, $y$, and $z$ direction, respectively. The discretization used $512\times256\times256$ grid points in the respective coordinate directions, with finite element order $p=4$, corresponding to a total of 33.6 million grid points. Isothermal walls were specified at $y=0$ and $y=2\delta_c$, while periodic boundaries were used in x- and z- directions. Starting from the center of the channel, the grid was progressively compressed in the y-direction to best capture the gradients at the wall, using
\begin{equation}
    y' = 2\pi\cdot\left(y-0.8\cdot\frac{\sin(2\pi y)}{2\pi}\right).
\end{equation}
The first grid point in the wall-normal direction was thus located at $y^+ = y u_\tau/\nu_w=0.25$, with kinematic viscosity $\nu_w$ at the wall and friction velocity $u_\tau=\sqrt{\tau_w/\rho_w}$ being defined by the shear stress $\tau_w$ and density $\rho_w$.
The physical lengths in the following discussion are normalized using the friction length scale $l^+=\nu_w/ u_\tau$, indicating those quantities using a superscript $^+$.
The D3Q45 discretization was used for the velocity space.

\begin{figure}[ht!]
    \centering
    \includegraphics[width=0.7\linewidth]{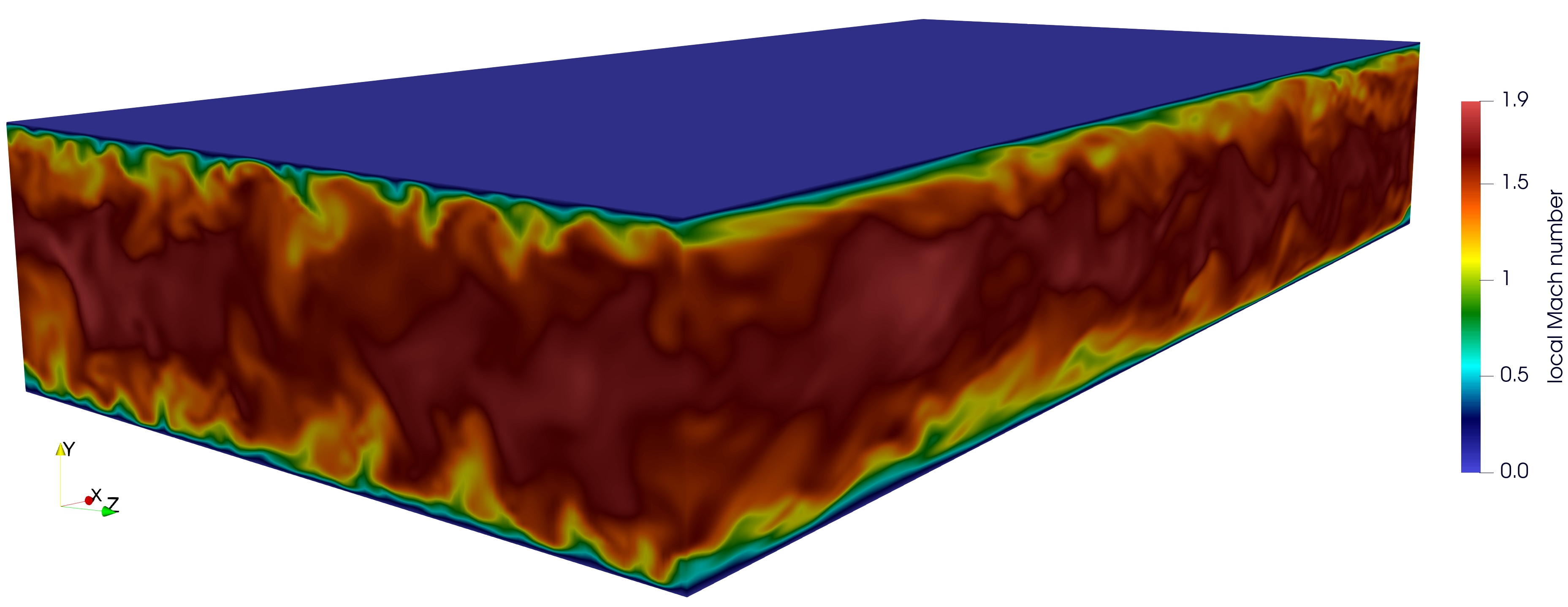}
    \caption{3D view of the supersonic channel, colored by the local Mach number}
    \label{fig:channel}
\end{figure}

A 3D view of the channel is shown in Fig.~\ref{fig:channel} and gives an impression of the instantaneous flow in the channel based on the local Mach number. It can be seen that a wide range of Mach numbers is realized: from subsonic regions in the viscous sublayer, buffer, and early logarithmic layer, to Mach numbers clearly exceeding the bulk centerline Mach number of $\mathrm{Ma}_b=1.5$ due to turbulent fluctuations.

To overcome friction and reach a statistically steady state, a volume force is usually added to the right-hand side of the momentum and energy equation \cite{Foysi:2004,Coleman:1995} in finite difference or volume codes, usually replacing the mean streamwise pressure gradient. Here, the pressure gradient in the streamwise direction was initially set using a volume force per unit volume, by approximating the acceleration via a velocity difference $\delta \boldsymbol{u}$ und time step $\delta_t$, such that
\begin{equation}
    F_x = \rho \cdot \frac{\delta \boldsymbol{u}}{\delta_t} 
    = \frac{\rho_w \mathrm{Re}_\tau^2 \mu_w^2}{\delta_c^3}
\end{equation}
and then subsequently adjusted using a PI controller, by fixing the averaged momentum flux. For this purpose, the \gls{lbm} forcing scheme of Kupershtokh was used \cite{Kuper:2009}, where the velocity difference $\delta \boldsymbol{u}$ is introduced into the distribution functions by means of a difference of two equilibrium functions after the collision step has been performed,
\begin{equation}
    \delta f_i = f^\mathrm{eq}_i(\rho,\boldsymbol{u}+\delta \boldsymbol{u},T) - f^\mathrm{eq}_i(\rho,\boldsymbol{u},T).
\end{equation} 
The walls of the channel need to be cooled to allow for supersonic flows. Therefore, an isothermal boundary condition was set by prescribing the wall temperature $T_w$, whereas the fluid heats up due to friction and compression. An equilibrium of the averaged temperature and flow quantities is reached after several through-flow times, determined by monitoring all flow quantities and statistics over time, as well as checking the stress balance. 

The test case is executed in three phases. It is initialized with Mach number $\mathrm{Ma}_b(t=0)=1.0$ on a coarse grid of $128\times64\times64$ grid points. Large mode, high-frequency perturbations were superimposed in all spatial directions and scaled by a parabolic profile in wall-normal direction. After a settling time of 100,000 time steps with $\delta_t = 0.0016 u_b/\delta_c$ to allow a transition to turbulence, the flow was accelerated to the target Mach number $\mathrm{Ma}_b(t\rightarrow\infty)=1.5$ via adjustment of the volume force. For the second phase, the distribution function values were transferred to the final grid in two steps by interpolation. After 200,000 time steps on the final grid, the final equilibrium was reached, and the acquisition and averaging of the variables were performed in a third phase over 80,000 time steps. The time-step size was $\delta_t = 0.0008 u_b/\delta_c$ on the final computational grid for approximately 5.1 through-flow times.

Due to the interpolation required in the \gls{sllbm} flow step, a minor mass loss of about $0.0001\%$ occurred in each time step. To keep the total mass constant over the complete simulation time, all distribution function values of the simulation were scaled by a factor $\rho_b(t=0) / \rho_b(t>0)$ every ten time steps.

\begin{figure}[t!]
    \centering
    \includegraphics[width=.78\textwidth]{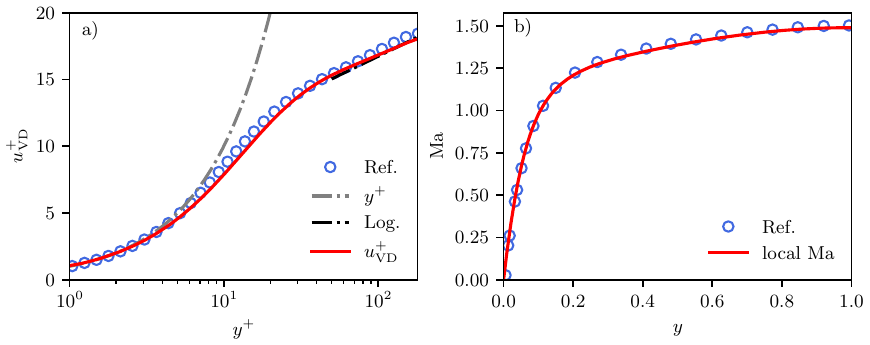}
    \vspace{-.2cm}
    \caption{a): Van-Driest transformed velocity $u^+_\mathrm{vD}$ over $y^+$. Reference from Ruby and Foysi \cite{Ruby:2022}. b): Averaged local Mach number over the course of half the channel height. Reference from Foysi \cite{Foysi:2004}}
    \label{fig:vanDriest}
\end{figure}

The Van Driest transformed velocity over $y^+$ and the mean Mach number profile are shown in Fig.~\ref{fig:vanDriest}. The Van Driest transformation $u^+_\mathrm{vD}$ takes into account the density variation perpendicular to the wall and scales the velocity as
\begin{equation}
u^+_\mathrm{vD} = \int_0^{u^+} \frac{\overline{\rho}}{\overline{\rho_w}} du^+.
\end{equation}

\noindent The scaling of $u_\mathrm{VD}^+=y^+$ within the viscous sublayer is well reproduced. Moreover, very good agreement with the results of Ruby and Foysi \cite{Ruby:2022} is shown in Fig.~\ref{fig:vanDriest} a).
As $y^+$ progresses, the logarithmic law of the wall $u^+(y^+)= \mathrm{ln}(y^+)/0.41 + 5.5$ is also captured by the Van Driest transformed velocity, showing the compressibility effects mainly to be variable density effects \cite{Foysi:2004}.  Fig.~\ref{fig:vanDriest} b) depicts the very good agreement of the local Mach number across the channel half-width, compared to data from Foysi \emph{et al.} \cite{Foysi:2004}.

\begin{figure}[b!]
    \centering
    \includegraphics[width=.78\textwidth]{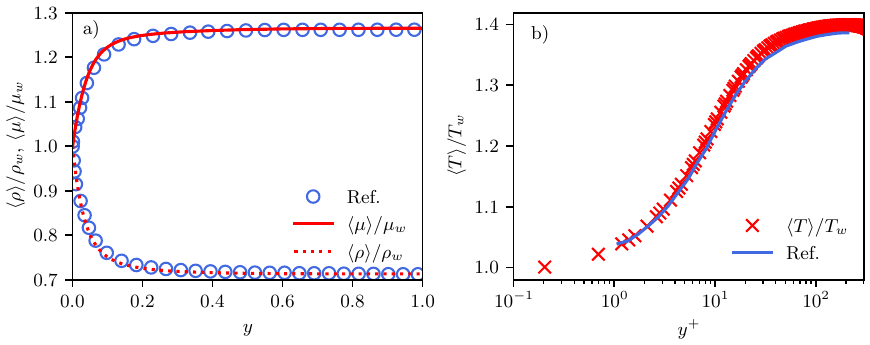}
    \vspace{-.3cm}
    \caption{a): Averaged dynamic viscosity and density normalized by their respective values at the wall over the course of half the channel height compared to Ruby and Foysi \cite{Ruby:2022}. b): Averaged temperature near the wall compared to Foysi \cite{Foysi:2004} values.}
    \label{fig:channel_T}
\end{figure}

Fig.~\ref{fig:channel_T} a) shows the temperature-dependent dynamic viscosity and density, each normalized by the wall values. Again, the values agree well with those of Ruby and Foysi \cite{Ruby:2022}. Different resolutions and numerical scheme orders lead to small temperature deviations, due to differences in numerical dissipation regarding both codes. Additionally, a closer look when using $y^+$-coordinates and a logarithmic axis (Fig.~\ref{fig:channel_T} b) shows that the temperature values increase slightly faster, here, even though the temperature in the middle of the channel matches the reference values. These small differences are attributable to the distinct numerical schemes and grid stretching utilized in this work, as Ruby \& Foysi \cite{Ruby:2022} use high-order methods.

\begin{figure}[ht!]
    \centering
    \includegraphics[width=.8\textwidth]{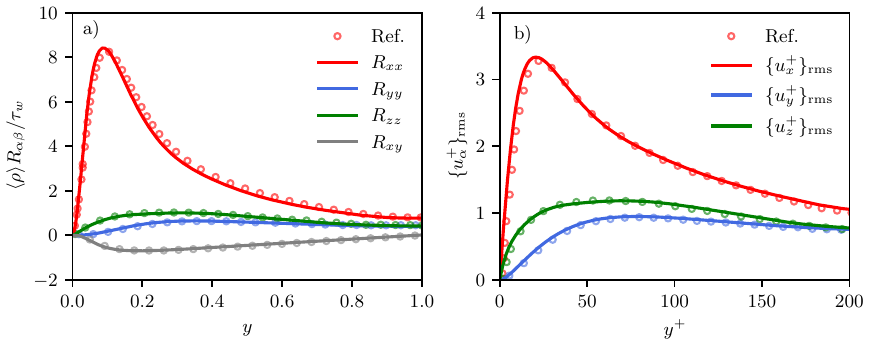}
    \vspace{-.3cm}
    \caption{a): Selected entries of the Reynolds stress tensor compared to the reference of Ruby and Foysi \cite{Ruby:2022}. b): \gls{rms} values of the normalized velocities $u_\alpha^+$ compared to the values of Foysi \cite{Foysi:2004}.}
    \label{fig:channel_stress_tensor}
\end{figure}

Finally, in Fig.~\ref{fig:channel_stress_tensor} a) and b), selected components of the Reynolds stress tensor are plotted in outer and inner coordinates using $y^+$, respectively. The Reynolds stress tensor $R_{ij}$ for compressible flows is usually given by
\begin{equation}\label{eq:rij}
   \overline{\rho} R_{ij} = \overline{\rho}\widetilde{u_i'' u_j''} = \overline{{\rho u_i'' u_j'' ~}}.
\end{equation}
Again, the reference values of Ruby and Foysi \cite{Ruby:2022} are reproduced by the \gls{sllbm} simulation, showing the capability of our compressible \gls{lbm} method to accurately simulate three-dimensional supersonic compressible turbulent flow.  
\subsection{Compressible 3D Mixing Layer}\label{sec:mixing}

\begin{figure}[b!]
\vspace{-.2cm}
\centering\includegraphics[width=0.5\linewidth]{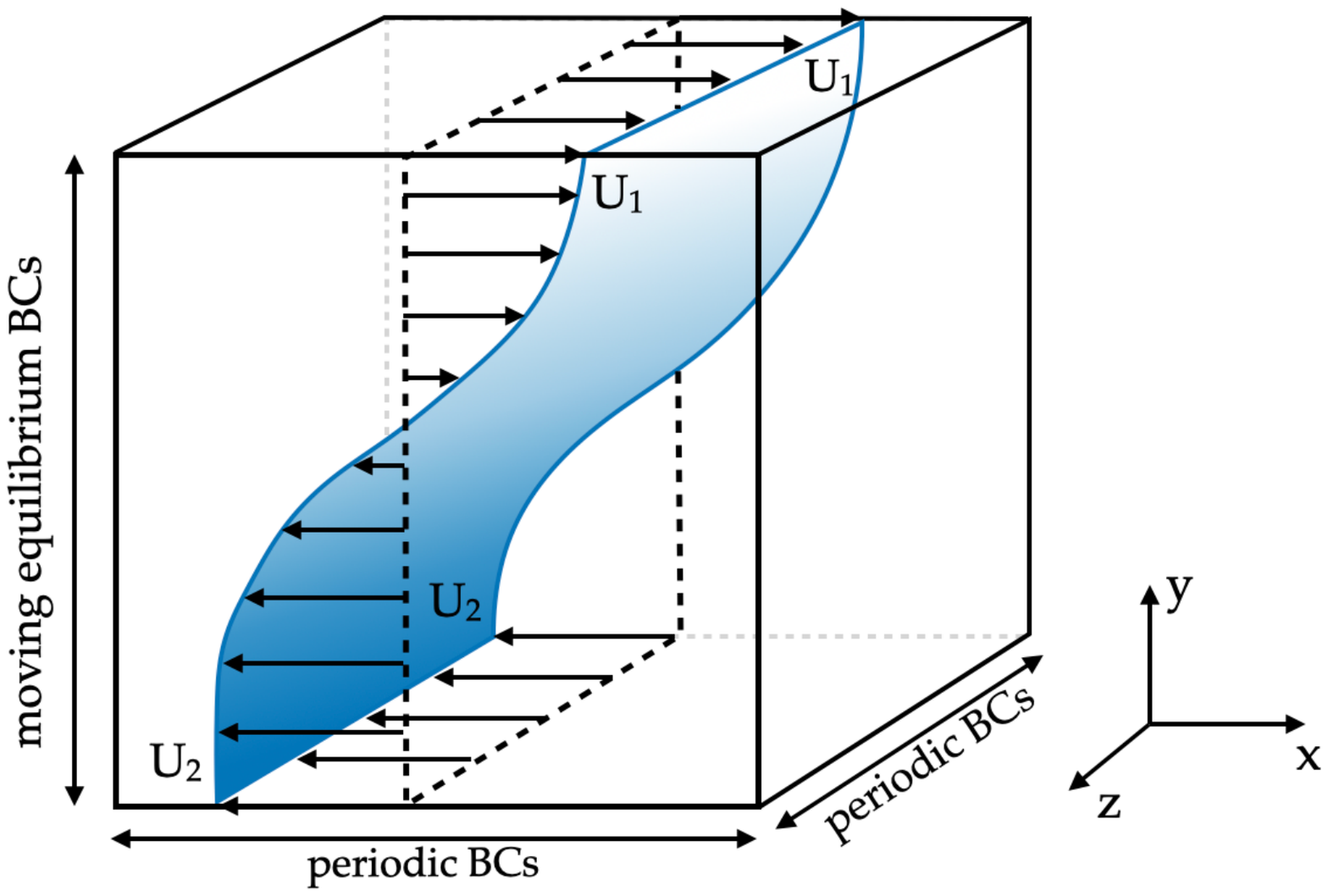}
    \vspace{-.3cm}
    \caption{Sketch of the domain and flow field for a temporal mixing layer, with $U_1 = -U_2$ and $\rho_1=\rho_2$}
    \label{fig:ux0}
\end{figure}

For a final test, compressible three-dimensional supersonic temporally developing turbulent mixing layers were simulated, for the first time using \gls{lbm}, to the best of our knowledge. This test serves to judge the ability of the present \gls{sllbm} code to capture the three-dimensional transition to turbulence by accurately predicting the instability and subsequent shear layer growth, as well as compressibility effects on the flow structure. The mixing layer is set up in a Cartesian coordinate system with streamwise and cross-stream direction $x=x_1$ and $z=x_3$, respectively, as well as shear direction $y=x_2$. Two streams with prescribed velocities ${U}_1 = -{U}_2$ and velocity difference $\Delta u_0 = U_1 - U_2$ are considered, as depicted in Fig. \ref{fig:ux0}.
The streamwise velocity is initialized as 
\begin{equation}
    u_{1}(x_2,t=0) = \frac{\Delta u_0}{2}\cdot\tanh\left(-\frac{x_2}{\delta_{\theta 0}}\right),
\end{equation}
with $\delta_{\theta 0}$ being the initial momentum thickness (see definition below), set as $\delta_{\theta 0}=0.093$.
The background density $\rho_0$ and temperature $T_0$ at $t=0$ were prescribed as unity, and the velocity difference as $\Delta u_0 = 2$.
Furthermore, the ratio of the specific heats was set to $\gamma=1.4$, the Prandtl number to $\mathrm{Pr}=0.71$, and the Reynolds number based on the initial momentum thickness and velocity difference to $\mathrm{Re}=800$. The initial Reynolds number based on the vorticity thickness $\delta_{\omega}$, to be defined below, was set to $\mathrm{Re}_{\omega 0}=\rho_0\mathrm{Re}\Delta u_0 \delta_{\omega0}= 610$. The characteristic Mach number usually used for describing such flows is the convective Mach number, defined as $\mathrm{Ma}=(U_1-U_2)/(c_1+c_2)$, using the speeds of sound $c_i$ for the two streams at $t=0$ and the velocity difference.
\begin{figure}[b!]
\centering
\includegraphics[height=7cm]{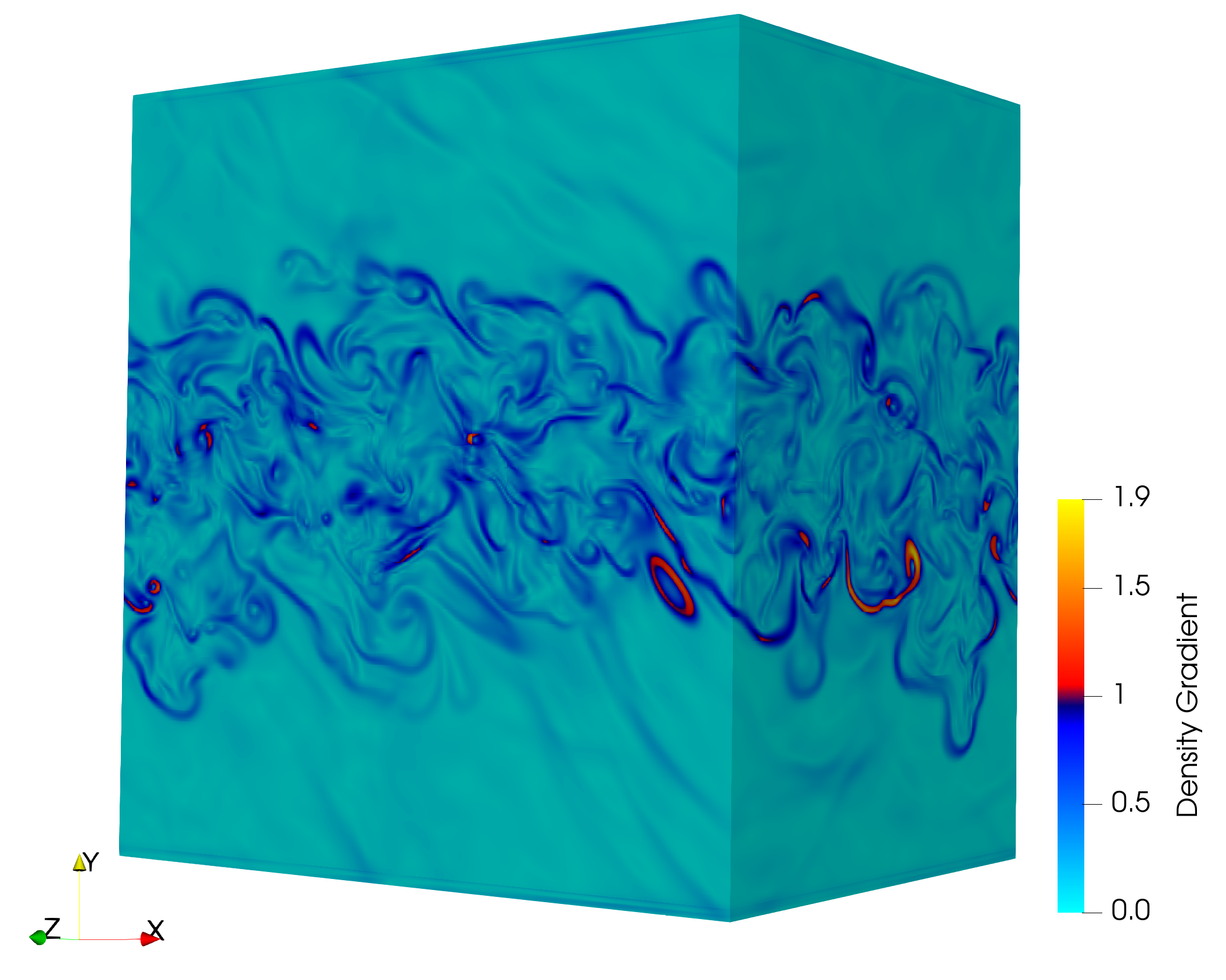}
\vspace{-.9cm}
 \caption{3D view of the compressible mixing layer at $\mathrm{Ma}=0.9$ and $\tau=1355$, colored by the density gradient.}
 \label{fig:mix_vis}
\end{figure}
\begin{table}[t!]
    \centering
    \caption{Grid size and resolution using a relative grid stretching of 10\% outside the central region in $x_2$-direction. Values are normalized by $\delta_{\theta 0}$. The grid resolution is $\Delta_i \in [\Delta_{i,\min}, \Delta _{i,\max}]$ due to the solver using the Gauß-Lobatto-Chebycheff distribution of integration points, with $i\in{x,y,z}$.}
   {\small \begin{tabular}{r|c|c|c}
         & $x$ & $y$ & $z$\\
         \hline
        $l_i$ & 160 & 176 & 98\\
        $N_i$ & 289 & 289 & 177\\
        $\Delta_i$ & [0.325, 0.786] & [0.325, 1.517] & [0.326, 0.787]
    \end{tabular}}
    \label{tab:mixdomain}
\end{table}
The domain size was prescribed as  
$l_x \times l_y \times l_z = 160\times176\times98$,  normalized by the initial shear layer thickness $\delta_{\theta0}$. The computational grid was stretched outside the shear region in $y$-direction, by using an equidistant grid in the central region $y\in[-64,64]$, connected to a parabolic function $y'=ay^2+by+c$,
coarsening the grid outside this region, acting as a sponge region close to the boundaries.
Tab.~\ref{tab:mixdomain} lists the minimum and maximum grid spacing and grid points. 
The discretization utilized $72\times72\times44$ elements of order $p=4$, resulting in 14.78 million grid points in total. The time-step sizes are given in Tab.~\ref{tab:mix_times}. The flow was discretized using the D3Q45 velocity space. The domain made use of periodic boundary conditions in the $x_1$ and $x_3$-directions and moving equilibrium boundary conditions with given density $\rho=\rho_0$, temperature $T=T_0$, and velocity $u_1=\pm\frac{\Delta u_0}{2}$ in $x_2$-direction. 
Tab.~\ref{tab:mix_times} furthermore shows the maximum simulation time $t_{max}$ and the computing time required on 32 nodes (see Sec.~\ref{sec:results} for details) for different $\mathrm{Ma}$. Additionally, the start and end times ($\tau_{min/max}$ with $\tau = t \Delta u_0/\delta_{\theta 0}$) describing the extent of the self-similar stage of the simulations, are listed.

\begin{table}[!t]
    \centering
    \caption{Minimum and maximum times $\tau_{min/max}=\{t \Delta u_0/\delta_{\theta0}\}_{min/max}$ indicating the extend of the self-similar region, maximum time of the simulation $t_{max}$, time-step size $\Delta t$ and the approximate required computation time $T[h]$ in hours for different $\mathrm{Ma}$.}
    {\small\begin{tabular}{r|c|c|c|c|c}
         & $\mathrm{Ma}=0.3$ & $\mathrm{Ma}=0.7$ & $\mathrm{Ma}=0.9$ & $\mathrm{Ma}=1.2$\\
        \hline
        $\tau_\mathrm{min}$ & 350 & 450 & 800 & 1400\\
        $\tau_\mathrm{max}$ & 550 & 700 & 1100 & 2000\\
        $t_\mathrm{max}$ & 25.11 & 32.55 & 51.15 & 93\\
        $\Delta t$ & 0.0022 & 0.0052 & 0.0066 & 0.0088\\
        $T[h]$ & 12 & 6.5 & 12 & 13.5\\
    \end{tabular}}
    \label{tab:mix_times}
\end{table}

Broadband disturbances are superimposed in all directions. They are obtained by proceeding similarly to \cite{foysiCompressibleMixingLayer2010}, creating a randomized velocity potential field $\mathbf{\Xi}_0$ with values equally distributed between 0 and 1. To restrict it to the initial shear layer thickness and mimic a prescribed velocity spectrum, the field was scaled in spectral space using the relation $\hat{\mathbf{\Xi}}(\mathbf{k})=\hat{\mathbf{\Xi}}_0(\mathbf{k})\cdot e^{-|\mathbf{k}|/24}$ first, then restricting it to the shear layer by multiplying the field in physical space by $\exp[-(x_2/2\delta_{\theta 0})^2]$. The velocity disturbance field is afterward obtained by taking the curl of the potential field via $\mathbf{u}=\nabla\times\mathbf{\Xi}$, thereby guaranteeing solenoidality. This method was demonstrated by Erlebacher \emph{et al.} \cite{erlebacher1990analysis}, too, to allow a fast transition to turbulence with limited initial acoustic waves. Fig. \ref{fig:mix_vis} depicts the structure of a developed turbulent mixing layer at $M_c=0.9$ by visualizing the density gradient, indicating localized regions of very steep density changes.

The temporal mixing layer approaches a self-similar stage after an initial transient. This has been confirmed in several experiments as well as numerical simulations for incompressible \cite{jones1971statistical,bell1990development,rogers1994direct} and compressible flows \cite{foysiCompressibleMixingLayer2010,papamoschou1988compressible,elliott1990compressibility,barre1994compressibility,chambres1998detailed,Vreman:1996,freund2000compressibility,pantanoStudyCompressibilityEffects2002}. The major variables, i.e., stress tensors, vorticity thickness, and momentum thickness, were reproduced using the \gls{sllbm}. They are briefly discussed here.

\begin{figure}[t!]
    \centering
        \centering
        \includegraphics[width=0.7\textwidth]{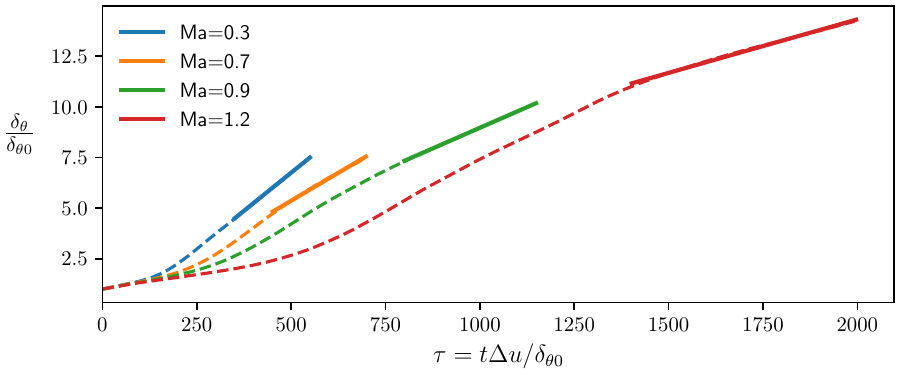}
        \vspace{-.2cm}
        \caption{Momentum thickness normalized by its initial value over normalized time for different Mach numbers. The linear growth observed within the self-similar region is indicated by superimposed straight lines.}
        \label{fig:deltaTheta_comparison}
        \vspace{0.0cm}
\end{figure}

The vorticity thickness $\delta_\omega$ and momentum thickness $\delta_\theta$ are calculated as 
\begin{equation}
    \delta_\omega = \frac{\Delta u_0}{\left( \partial \overline{u}_1 / \partial x_2 \right)_{\max}}  
    \vspace{-.1cm}
\end{equation}
and
\vspace{-.1cm}
\begin{equation}
    \delta_\theta = \frac{1}{\rho_0 \Delta u_0^2} \int_{-\infty}^{\infty} \overline{\rho} \left(\frac{1}{2}\Delta u_0 - \tilde{u}_1\right) \left(\frac{1}{2}\Delta u_0 + \tilde{u}_1\right) dx_2,
\end{equation}
respectively. 
Within the self-similar stage, the momentum and vorticity thickness are approximately related as $\delta_\omega\approx5\delta_\theta$. The momentum thickness is plotted in Fig.~\ref{fig:deltaTheta_comparison}, normalized by its initial value as a function of the normalized time variable $\tau = t \Delta u_0 / \delta_{\theta 0}$. The plot indicates a decrease in the growth rate within the self-similar region, depicted using straight lines where self-similar linear growth occurs. Additionally, the delayed growth of the compressible mixing layer with increasing convective Mach number \cite{pantanoStudyCompressibilityEffects2002} is clearly visible.
\begin{figure}[b!]
        \centering        \includegraphics[width=0.7\textwidth]{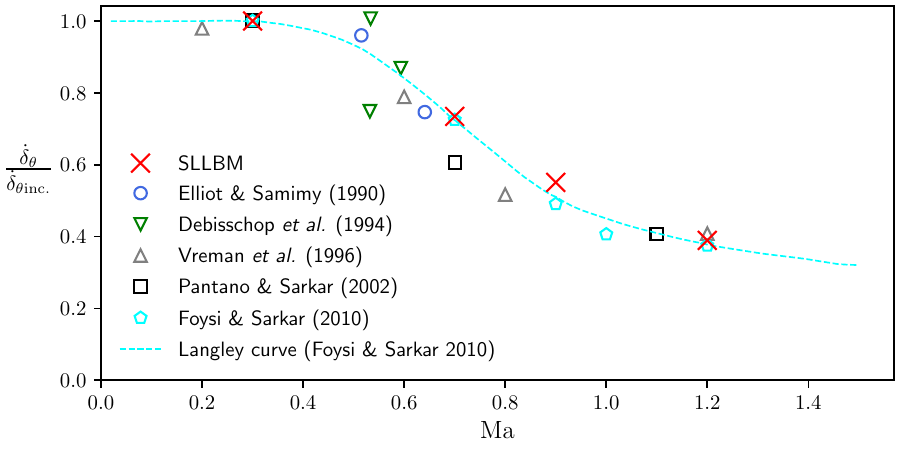}
        \vspace{-.15cm}
        \caption{Shear layer growth rate normalized by its incompressible value, $\dot{\delta}_{\theta\mathrm{inc.}}$, as a function of the convective Mach number, $\mathrm{Ma}$. The results are compared with experimental data 
        from Elliot \& Samimy \cite{elliott1990compressibility}, Debisschop, Chamres, \& Bonnet \cite{debisschopVelocityFieldCharacteristics1994}, and Vreman, Sandham, \& Luo \cite{Vreman:1996}, as well as simulation data from Foysi \& Sarkar \cite{foysiCompressibleMixingLayer2010} and Pantano \& Sarkar \cite{pantanoStudyCompressibilityEffects2002}.}
        \label{fig:deltaThetaDot_Ma}
\end{figure}
\noindent For the quasi-incompressible case, the growth rate of the \gls{sllbm} simulation was determined as $\dot{\delta_\theta} \approx 0.0155$, comparing favorably with simulations from \cite{pantanoStudyCompressibilityEffects2002,foysiCompressibleMixingLayer2010}. 
Fig.~\ref{fig:deltaThetaDot_Ma} 
shows a comparison of the momentum thickness normalized with the incompressible value and the growth rates plotted over the convective Mach number $\mathrm{Ma}$, respectively. The absolute values can be furthermore found in Tab.~\ref{tab:mix_comp}. The growth rates were determined within the phase of the self-similar temporal development of the mixing layer, indicated by the straight lines in  Fig.~\ref{fig:deltaTheta_comparison}. Fig.~\ref{fig:deltaThetaDot_Ma} compares the data with experimental results and a data-fit referred to as the ’Langley Experimental Curve’. This curve comprises a compilation of measurements of air–air shear layer experiments \cite{foysiCompressibleMixingLayer2010}. The results achieved with the \gls{sllbm} are in excellent agreement with the results in the literature. The growth-rate reduction with increasing convective Mach number is captured nicely, too. For large convective Mach numbers shocklets, as described by Fu \& Li \cite{fu2006numerical} and Foysi \& Sarkar \cite{foysiCompressibleMixingLayer2010}, were temporally observed and stably resolved.
The growth-rate reduction indicates a reduced mixing efficiency due to compressibility effects. This was discussed in various publications investigating (a) the linear stability of mixing layers and reduced growth rate of stability waves with increasing convective Mach numbers, (b) the breakdown of communication between different flow regions, (c) and reduced pressure fluctuations acting via a reduction of the pressure-strain correlation \citep{foysiCompressibleMixingLayer2010,papamoschou1988compressible,freund2000compressibility,pantanoStudyCompressibilityEffects2002,Sandham:1990,Papamoschou:1993}. A nice overview can be found in Kwok \cite{Kwok:2002}.

\begin{table}[t!]
    \centering
    \caption{Comparison of the momentum thickness growth rates and normalized peak turbulent intensities, averaged within the self-similar range\vspace{-0.2cm}}
    \label{tab:mix_comp}
    \begin{tabular}{r|c|c|c|c|c|c|c|c|}
        & $\mathrm{Ma}$ & $\dot{\delta_\theta}$ & $\frac{\sqrt{R_{11_\mathrm{.max}}}}{\Delta u_0}$ & $\frac{\sqrt{R_{22\mathrm{.max}}}}{\Delta u_0}$ & $\frac{\sqrt{R_{33\mathrm{.max}}}}{\Delta u_0}$ & $\frac{\sqrt{R_{12\mathrm{.max}}}}{\Delta u_0}$ & $\sqrt{\frac{R_{22\mathrm{.max}}}{R_{11\mathrm{.max}}}}$ & $\sqrt{\frac{R_{12\mathrm{.max}}}{R_{11\mathrm{.max}}}}$ \\
        \hline
        \gls{sllbm} & 0.3 & 0.0147 & 0.173 & 0.129 & 0.154 & 0.108 & 0.749 & 0.627 \\
        \gls{sllbm} & 0.7 & 0.0108 & 0.165 & 0.111 & 0.135 & 0.095 & 0.675 & 0.575 \\
        \gls{sllbm} & 0.9 & 0.0081 & 0.144 & 0.100 & 0.121 & 0.086 & 0.694 & 0.595 \\
        \gls{sllbm} & 1.2 & 0.0057 & 0.137 & 0.094 & 0.114 & 0.078 & 0.690 & 0.571 \\
    \end{tabular}
    \vspace{-0.2cm}
\end{table}

\begin{figure}[bh!]
\vspace{-0.cm}
    \centering
    \includegraphics[width=0.95\textwidth]{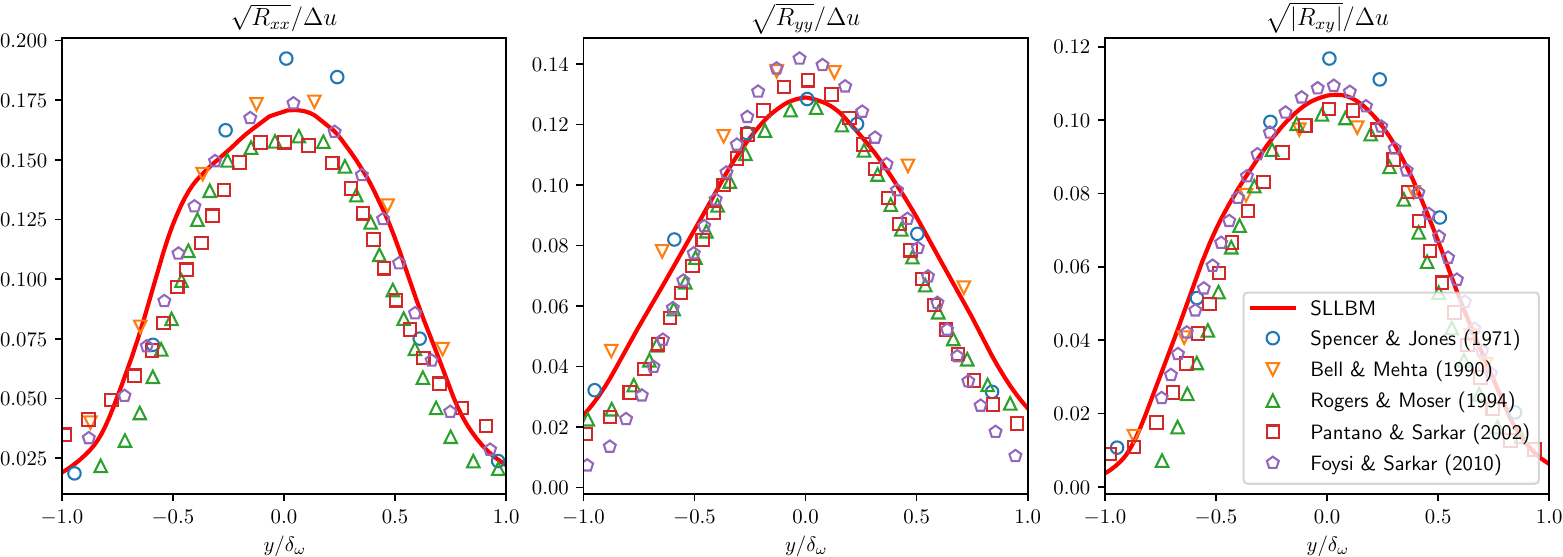}
    \caption{Profiles for the \gls{rms} values of the velocity fluctuations at $\mathrm{Ma}=0.3$, compared to the values of Foysi \& Sarkar \cite{foysiCompressibleMixingLayer2010}, Pantano \& Sarkar \cite{pantanoStudyCompressibilityEffects2002}, Bell \& Mehta \cite{bell1990development}, Spencer \& Jones \cite{spencer1971statistical}, and Rogers \& Moser \cite{rogers1994direct}.}
    \label{fig:RijRefMa03}
    \vspace{0.3cm}
\end{figure}

\begin{figure}[h!]
    \includegraphics[width=0.95\textwidth]{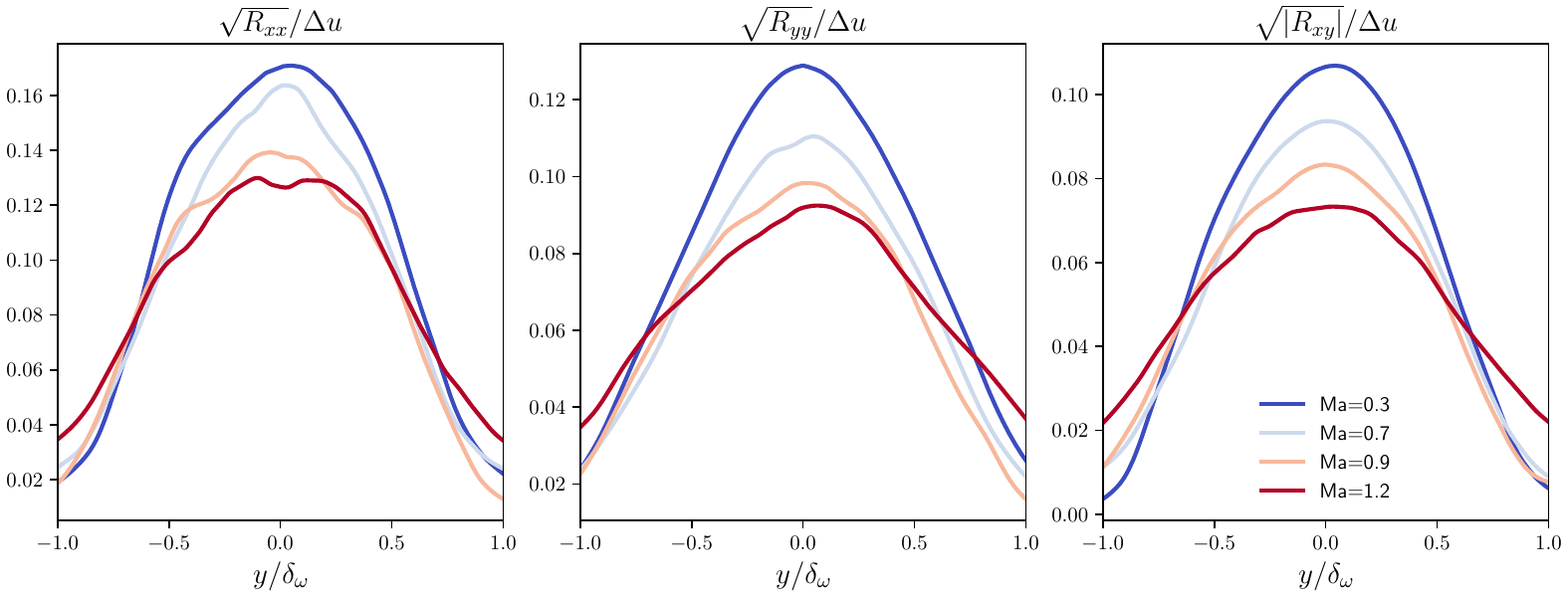}
    \caption{Comparison of the \gls{rms} velocity fluctuations for different Mach numbers, demonstrating the impact of compressibility.}
    \label{fig:RijMaComp}
    \vspace{0.3cm}
\end{figure}

\begin{figure}[h!]
    \includegraphics[width={0.95\textwidth}]{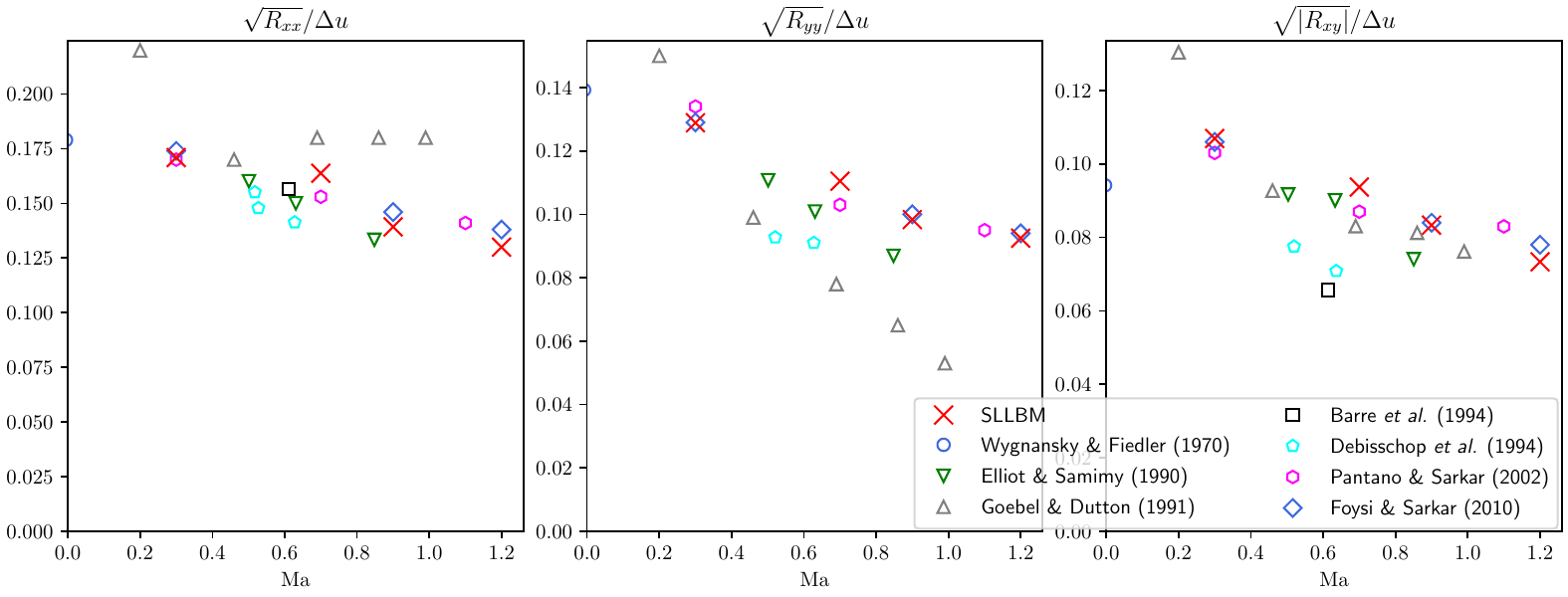}
    \caption{Comparison of the peak \gls{rms} velocity fluctuations across Mach numbers, compared to the experimental results of Wygnansky \& Fiedler \cite{wygnanski1970two}, Elliot \& Samimy \cite{elliott1990compressibility}, Goebel \& Dutton \cite{goebelExperimentalStudyCompressible1991}, Barre, Quine \& Dussauge \cite{barre1994compressibility}, and Debisschop, Chamres, \& Bonnet \cite{debisschopVelocityFieldCharacteristics1994}, and to the numerical results of Pantano \& Sarkar \cite{pantanoStudyCompressibilityEffects2002} and Foysi \& Sarkar \cite{foysiCompressibleMixingLayer2010}.}
    \label{fig:RijRefPeaks}
\end{figure}

The turbulent stress tensor $\bar\rho R_{ij}=\bar\rho \widetilde{ u_i'' u_j''}=\overline{\rho u_i'' u_j''}$ (Eq. \ref{eq:rij}) is a key quantity in analyzing the turbulence structure and the target of various modeling procedures. 
A selection of the tensor's elements was averaged over the self-similar stage and is compared to literature data in Fig. \ref{fig:RijRefMa03}, peak values are furthermore listed in Tab.~\ref{tab:mix_comp}.
The results show excellent agreement with experimental data and simulation results for $\mathrm{Ma} =0.3$ (Fig. \ref{fig:RijRefMa03}). 
Fig. \ref{fig:RijMaComp} shows the typical effect of compressibility, reducing the Reynolds stresses in magnitude, having been linked due to a reduction in pressure strain correlation, as analyzed in detail in Foysi \& Sarkar \cite{foysiCompressibleMixingLayer2010}.
The \gls{sllbm} shows slightly larger peak values and broader profiles compared to the \gls{dns} of Pantano \& Sarkar \cite{pantanoStudyCompressibilityEffects2002}, more in line with the results in Foysi \& Sarkar \cite{foysiCompressibleMixingLayer2010}. 
When further compared with various experimental and numerical results from the literature, the \gls{sllbm} is seen to agree with the literature results very well and captures the trends and magnitude of the peak turbulent stresses within the self-similar region for various convective Mach numbers, as shown in Fig. \ref{fig:RijRefPeaks}.
Differences can be furthermore explained by different initial perturbations, boundary conditions, domain sizes, effects of wind tunnel walls, resolution, or numerical schemes. 


\section{Conclusion}\label{sec:conclusion}

This paper demonstrates the application of the \acrlong{sllbm} to various compressible flow cases, including three-dimensional supersonic flow around spheres, turbulent fully developed three-dimensional compressible supersonic channel flow, and temporal mixing layers.

Contrary to low Mach number, quasi-incompressible flow, there is still considerable ambiguity regarding the most promising approaches to simulate flows at high Mach numbers, including shocks, with \gls{lbm}. As discussed above, extra stabilizing measures and very large velocity discretizations are often required, making simulations of high Mach number flows in three dimensions using \gls{lbm} costly.

Here, we utilize the \gls{sllbm} from Kr\"amer \emph{et al.} \cite{Kraemer:2017}, adjusted to be suitable to simulate high Mach number flows in \cite{Wilde2020,Wilde2021,Wilde2021a}, making use of reduced off-lattice velocity discretizations and the \gls{sllbm} stability advantages allowing separate time-step and grid size adjustments, as well as unstructured grids if required. The reduced D2Q19, D2Q25, and D3Q45 discretizations, being possible due to off-lattice points as detailed in \cite{Wilde2021}, allow for efficient simulation of supersonic two- and three-dimensional flow cases. 

To the best of our knowledge, turbulent fully developed three-dimensional compressible supersonic channel flow and temporal mixing layers were simulated using \gls{lbm} for the first time. The forcing scheme of Kupershtokh \emph{et al.} \cite{Kuper:2009} was used to generate the volume force necessary to replace the mean streamwise gradient and drive the channel flow. Suitable isothermal boundary conditions were implemented to cool the walls to allow supersonic flow. Additionally, the \gls{sllbm} allows for easy use of stretched grids, similar to those used in standard finite-difference solvers in the literature, suitable to tailor the flow to regions of large gradients. The results were in excellent agreement with results from the literature using standard finite difference or discontinuous Galerking solvers, demonstrating the capabilities and accuracy of the present compressible \gls{sllbm}.\\

\noindent {\bf Acknowledgement}
The authors thank the Deutsche Forschungsgemeinschaft (DFG) for providing funds via the DFG project FO 674/17-1. The authors thank the computer center of the Universit\"at Siegen (ZIMT) for granting the computational resources (OMNI cluster).


 \bibliographystyle{elsarticle-num} 
 \bibliography{mybib}





\end{document}